\RequirePackage{amsthm}

\documentclass[iicol,sn-mathphys-num,pdflatex]{sn-jnl}% Math and Physical Sciences Numbered Reference Style 
\usepackage{graphicx}%
\usepackage{multirow}%
\usepackage{amsmath,amssymb,amsfonts}%
\usepackage{mathrsfs}%
\usepackage[title]{appendix}%
\usepackage{xcolor}%
\usepackage{textcomp}%
\usepackage{manyfoot}%
\usepackage{booktabs}%
\usepackage{algorithm}%
\usepackage{algorithmicx}%
\usepackage{algpseudocode}%
\usepackage{listings}%
\usepackage{pifont}%
\usepackage{mathrsfs}
\usepackage{anyfontsize}
\usepackage{bookmark}
\theoremstyle{thmstyleone}%
\theoremstyle{thmstyletwo}%

\theoremstyle{thmstylethree}%

\renewcommand{\paragraph}[1]{\vspace{1em}\noindent\textbf{#1}\par}
\begin{document}

%\title[Article Title]{Three dimensional reconstructions of capillary flows and objects using Spatio-Temporal Phase Shifting Profilometry}

\title[Article Title]{Free surface topography of capillary flows using spatio-temporal phase shifting profilometry}


\author*[1]{\fnm{Hélie} \sur{de Miramon}}
\email{heliedemiramon@gmail.com}

\author[1]{\fnm{Wladimir} \sur{Sarlin}}
\email{wladimirsarlin@outlook.com}

\author[2]{\fnm{Axel} \sur{Huerre}}
\email{axel.huerre@u-paris.fr}

\author[3]{\fnm{Pablo} \sur{Cobelli}}
\email{cobelli@df.uba.ar}

\author[4]{\fnm{Thomas} \sur{Séon}}
\email{thomas.seon@cnrs.fr}

\author[1]{\fnm{Christophe} \sur{Josserand}}\email{christophe.josserand@polytechnique.edu}

\affil*[1]{\orgdiv{Laboratoire d'Hydrodynamique, CNRS, École polytechnique}, \orgaddress{\street{Institut Polytechnique de Paris}, \city{Palaiseau}, \postcode{91120}, \country{France}}}

\affil[2]{\orgdiv{Laboratoire Matière et Systèmes Complexes (MSC)}, \orgname{Université Paris Cité, CNRS, UMR 7057}, \city{Paris}, \postcode{75013}, \country{France}}

\affil[3]{\orgdiv{Universidad de Buenos Aires, Facultad de Ciencias Exactas y Naturales}, \orgname{Departamento de Física, IFIBA, CONICET, Ciudad Universitaria}, \city{Buenos Aires}, \postcode{1428}, \country{Argentina}}

\affil[4]{\orgdiv{Institut Franco-Argentin de Dynamique des Fluides pour l’Environnement (IFADyFE)}, \orgname{CNRS (IRL 2027), Universidad de Buenos Aires, CONICET}, \orgaddress{\city{Buenos Aires}, \postcode{1428}, \country{Argentina}}}

\abstract{We present a novel experimental technique for characterizing the free surface of capillary flows using the Spatio-Temporal Phase Shifting Profilometry (ST-PSP) method. This study specifically addresses various regimes of capillary flows over inclined surfaces, including drops, rivulets, meanders, and braided films. The technique is explained step by step with a detailed discussion of the calibration process, which is carried out on a solid wedge to determine the optical distances required for the phase-to-height relationship. In addition, the minimal dye concentration for accurately reconstructing the free surface of a dyed water flow is investigated. The ST-PSP method is then applied to profile different liquid flows, achieving large signal-to-noise ratios in all experiments. Notably, the analysis of a sessile droplet shows excellent agreement between the ST-PSP results and side-view visualizations, as demonstrated by the precise recovery of its apparent contact angle. Moreover, free surface reconstructions of rivulet flows align well with previous theoretical predictions. These findings suggest that the ST-PSP method is highly effective for obtaining precise height maps of capillary flows, offering a valuable tool for future validation of theoretical models.
}

\keywords{Phase shifting profilometry, Three dimensional reconstruction, Capillary flows, Contact line}



\maketitle

\clearpage

\section{Introduction}\label{sec1}

The study of capillary flows involving a contact line with a solid substrate reveals a rich interplay between inertia, surface tension, viscosity and wettability. Such liquid films appear in a wide range of applications, including coating processes \cite{2002_pasandideh-fard,2006_ristenpart}, inkjet printing \cite{2022_lohse}, soil erosion \cite{2015_zhao}, surface cleaning \cite{2016_aouad}, and capillary flow freezing \cite{2024_huerre}. Over the years, various flow regimes have been identified, encompassing deposited, impacting, or dripping droplets \cite{2004_biance,2016_josserand}, straight rivulets \cite{1966_towell,1995_duffy,2013_isoz,2022_monier}, stable or unstable meanders \cite{2006b_le_grand-piteira,2012_couvreur}, and braided films \cite{2005_mertens,2016_aouad,2018_grivel,2023_brient}. Despite their prevalence in Nature or in industry, the analytical modeling of these flows remains incomplete, largely due to strong assumptions often made about the flow dynamics, the free surface topography, or the fluid’s wettability \cite{1966_towell,1995_duffy,2005_mertens,2016_aouad}. Furthermore, experimental measurements of free surface deformations remain scarce, as they are notoriously difficult to obtain with high precision.

While pointwise measurements can be performed using laser profilometry \cite{2022_monier} or confocal chromatic imaging \cite{2005_lel}, these techniques are not well-suited for complex flows such as meanders or braided films and often lack accuracy near the contact line. Alternative methods, such as synthetic Schlieren \cite{2009_moisy}, allow for time-resolved measurements of the free surface of transparent liquids, but their applicability is limited to small curvatures and shallow depths due to the formation of caustics. A promising approach to overcoming these limitations is the adaptation of digital fringe projection techniques, initially developed for three-dimensional (3D) solid scanning, to the study of fluid flows. \citet{1983_takeda} introduced Fourier Transform Profilometry (FTP) as an initial approach, in which sinusoidal intensity fringes are projected onto a reference plane using a digital projector. When an object is placed on this reference plane, the fringes are distorted due to variations in the surface shape. By capturing images before and after the object of interest is introduced, the corresponding reference and deformed phase fields, respectively $\phi^{ref}(x,y)$ and $\phi(x,y)$, can be extracted through Fourier analysis, with $(x,y)$ denoting the spatial coordinates in the image. The phase difference, $\Delta \phi(x,y) = \phi(x,y)-\phi^{ref}(x,y)$, is then related to the height map using optical geometry principles \cite{2009_maurel}. A key advantage of FTP lies in its ability to provide time-resolved three dimensional reconstructions, as a single image is enough to extract the phase information. Consequently, FTP has been widely employed in the past two decades to investigate impacting drops \cite{2012_lagubeau}, liquid films on inclined surfaces \cite{2015_hu,2018_grivel}, surface waves \cite{2009_cobelli,2012_przadka}, and wave turbulence \cite{2022_falcon}. However, a major limitation of FTP remains its poor accuracy in detecting steep slopes.

\begin{figure*}
    \centering
    \includegraphics[width=\linewidth]{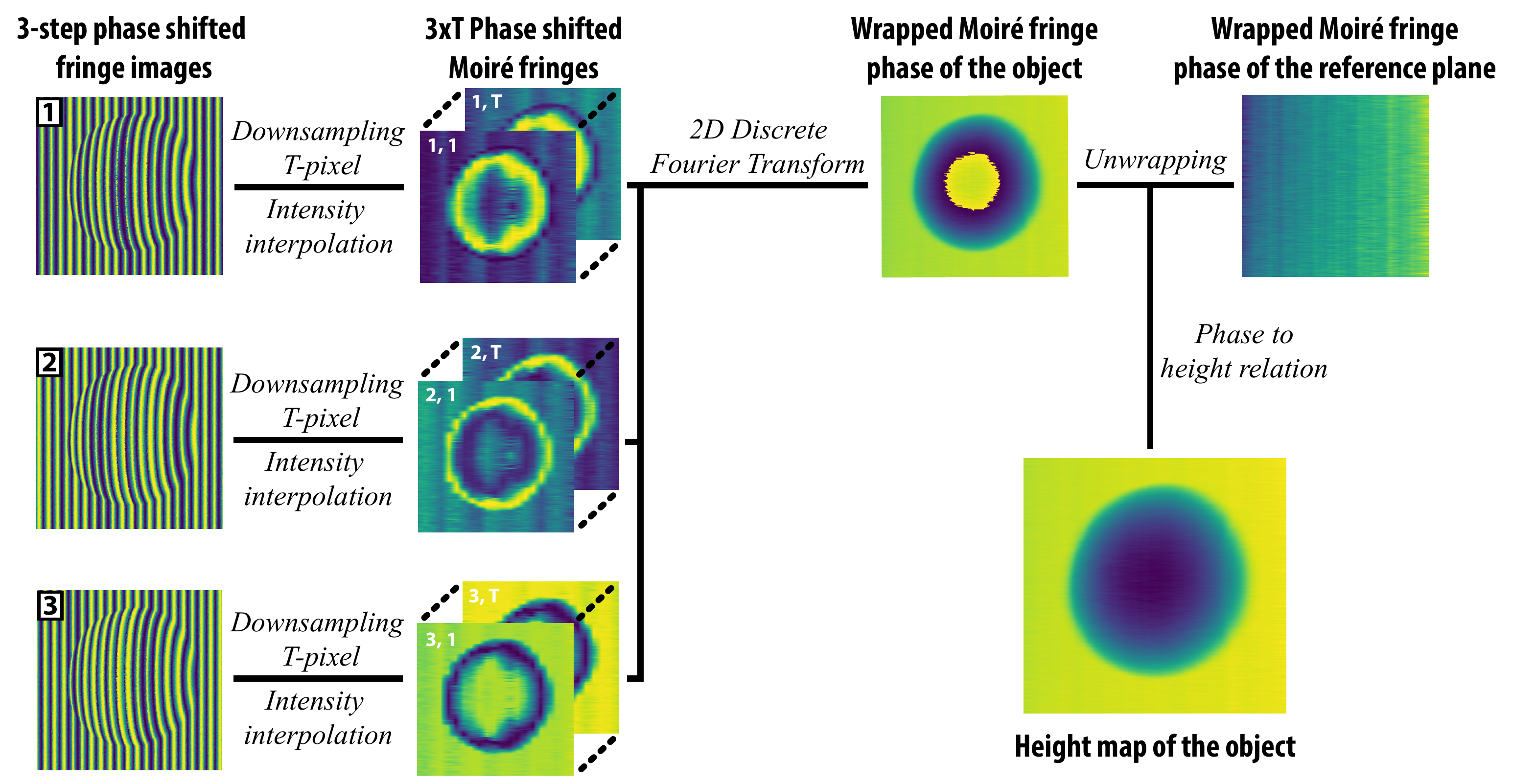}
    \caption{Schematic of the ST-PSP phase extraction applied to a spherical object, as described by \citet{2019_ri}. The unwrapping method relies on the algorithm developed by \citet{2002_Herraez}, while the phase-to-height relation is based on the theoretical model introduced by \citet{2009_maurel}.}
    \label{fig_principe_STPSP}
\end{figure*}

To address this long-standing challenge, \citet{1984_srinivasan} introduced the phase shifting profilometry (PSP) method, in which $N$ phase-shifted sinusoidal fringes ($N \geqslant 3$) are projected onto both the reference and the deformed planes, the latter resulting from the presence of the object. By performing algebraic operations on the $N$ phase-shifted images, one can determine $\phi^{ref}(x,y)$, $\phi(x,y)$, and consequently $\Delta \phi(x,y)$. In their study, the authors demonstrated that this approach enables high-resolution surface profiling of solid objects and is capable of handling steep slopes. Since then, numerous studies have significantly enhanced the accuracy and robustness of the PSP method, as highlighted in the review by \citet{2018_zuo}.

To further refine the detection of small deformations in solids, \citet{2010_ri} developed a variant of this approach, called the Sampling Moiré (SM) method. This method, based on single-shot phase analysis, uses a different image processing technique involving down-sampling and intensity interpolation. The authors reported highly precise measurements of small displacements in loaded steel beams, achieving an average error of only a few micrometers. More recently, \citet{2019_ri} introduced a spatio-temporal phase shifting profilometry (ST-PSP) method, combining both PSP and SM algorithms. This hybrid approach takes advantage of the strengths of each technique - such as a reduced sensitivity to noise, intensity saturation effects, and projection non-linearity - enabling highly accurate and robust three dimensional reconstructions of complex solid surfaces with significant variations in slope. Figure \ref{fig_principe_STPSP} illustrates the ST-PSP principle, combined with an unwrapping algorithm \cite{2002_Herraez} and a phase-to-height conversion \cite{2009_maurel}, to reconstruct the three dimensional height map of a spherical object. In this example, $N=3$ images were used, with a phase shift of $2\pi/3$ between consecutive projected gratings, as shown in the left column of the figure.

However, PSP, SM, and ST-PSP methods have never been used to investigate the free surface of flowing liquids, unlike synthetic Schlieren or FTP techniques. When applied to fluids, additional challenges arise, such as curved interfaces, temporal flow variations, and the need to maintain a sufficient dye concentration to ensure high light diffusivity at the flow free surface. In this study, the ST-PSP method is successfully applied for the first time to a dyed fluid interface, enabling the three-dimensional mapping of stationary or quasi-stationary capillary flows with a contact line. This, in turn, allows us for precise measurements of macroscopic contact angles along all edges of the flow free surface.

Section \ref{sec2} first reviews the principles of ST-PSP and the phase-to-height relation, before detailing the experimental setup used in the present investigation. A thorough description of the system calibration and parameter optimization process is also provided. Section \ref{sec3} then presents results for various capillary flow configurations involving a contact line: a sessile droplet, steady straight rivulets, a stationary meander, and a braided film. The method yields highly accurate measurements and successfully captures macroscopic contact angles. Finally, Section \ref{sec4} provides conclusions and perspectives for future work.

\section{Materials and methods}\label{sec2}

\subsection{Principles of the ST-PSP method}\label{subsec2.1}

The fundamental principle of the profilometry methods discussed in this study is as follows: a sinusoidal fringe pattern is projected onto an object using a video projector, and the resultant deformations of the fringes are captured by a camera. The analysis of fringe deformations allows for the determination of the phase $\phi$ of the image, which can then be correlated with the height map of the object.

%In the case of spatio-temporal phase shifting profilometry (ST-PSP), two distinct methods are combined: Phase shifting profilometry (PSP) \cite{2012_ri,2018_zuo} and sampling Moiré (SM) \cite{2010_ri}. 

The ST-PSP technique combines two distinct methods: PSP \cite{2012_ri,2018_zuo} and SM \cite{2010_ri}. In the first one, $N \geqslant 3$ phase-shifted images ($2 \pi/N$ shift between fringe patterns) are used in order to extract the image phase, that is related to the object height \cite{2009_maurel}. The SM method requires only one image of the deformed object as it relies on spatial interpolation \cite{2010_ri,2022_wang}.

\paragraph{PSP principle}
\noindent For the $N$-step PSP method, the general intensity profile of the $n$-th image ($n \in [0,N-1]$) is:

\begin{equation}
    \label{def_intensity_PSP}
    \begin{aligned}
        I_n(x,y) &= A(x,y) + B(x,y) \cos \biggl( \omega_0 x + \phi'(x,y) \\
        &  + \frac{2 n \pi}{N} \biggr) \\
        &= A(x,y) + B(x,y) \cos \left( \phi(x,y) + \frac{2 n \pi}{N} \right),
    \end{aligned}
\end{equation}

\noindent where $x$ and $y$ are the spatial coordinates of a point on the image, $A(x,y)$ and $B(x,y)$ are the background intensity and contrast of the image, respectively, and $\omega_0$ is the spatial frequency of the original fringe pattern on the reference plane. $\phi'(x,y)$ is the phase induced by the presence of the object and $\phi(x,y)$ is the overall phase of the image that can be extracted by performing a one dimensional discrete Fourier transform:

\begin{equation}
    \label{def_phi}
    \phi(x,y) = \arg \left\{ \sum_{n=0}^{N-1} I_n(x,y) e^{-\frac{2 i \pi n}{N}} \right\},
\end{equation}

\noindent with $i$ the imaginary number such that $i^2=-1$. Since $\phi$ is wrapped in the $[-\pi,\pi]$ interval, the discontinuous final phase map needs to be unwrapped. For this purpose the two-dimensional unwrapping algorithm developed by \citet{2002_Herraez} is used. 

Increasing the number $N$ of projected fringes can slightly improve the precision of the reconstructions. However, 3-step and 4-step reconstructions remain the most commonly used due to their experimental simplicity \cite{2018_zuo}.

\paragraph{SM principle}
\noindent On the other hand, the SM technique only requires one image, with an intensity profile $I_0(x,y)$; it then consists in the down sampling of that image with a period $T$, chosen as the integer closest to $\lambda_0 \equiv 2 \pi / \omega_0$ after scaling by the pixel ratio. After the interpolation of the intensity profiles, a number $T$ of phase-shifted Moiré fringe patterns of low frequency can be extrapolated, with the $t$-th image having an intensity $I_t$ such that

\begin{equation}
    \label{def_intensity_SM}
    \begin{aligned}
        I_t(x,y) &= A(x,y) + B(x,y) \cos \left( 2 \pi \left( \frac{1}{\lambda_0}-\frac{1}{T} \right) x \right. \\
        & \left. + \ \phi'(x,y) + \frac{2 t \pi}{T} \right) \\
        &= A(x,y) + B(x,y) \cos \left( \psi(x,y) + \frac{2 t \pi}{T} \right),
    \end{aligned}
\end{equation}

\noindent with $\psi(x,y)$ the phase that can then be extracted through a similar procedure as for the PSP method ($\phi(x,y)$, $I_n$ and $N$ would have to be replaced by $\psi(x,y)$, $I_t$ and $T$ respectively in equation \eqref{def_phi}). After a two-dimensional unwrapping of $\psi$, $\phi(x,y)$ is then determined by adding the sampling phase to $\psi(x,y)$:

\begin{equation}
    \label{def_relation_psi_phi}
    \phi(x,y) = \psi(x,y)+\frac{2\pi x}{T}.
\end{equation}

\paragraph{ST-PSP principle}
\noindent Since the phase information at each pixel can be extrapolated either temporally (PSP) or spatially (SM), the ST-PSP method combines both approaches to improve the accuracy and robustness of phase determination \cite{2019_ri}. This is achieved by processing a set of $N$ phase-shifted images and performing down-sampling and interpolation across all $N$ images. Using the $N$ phase-shifted images, the background intensity map $A$ and the contrast map $B$ can be determined, allowing for the normalization of intensity profiles. Subsequently, down-sampling and interpolation are applied to each image, resulting in a set of $N \times T$ images, with the $n$-th and $t$-th raw and normalized intensities $I_{n,t}$ and $\overline{I_{n,t}}$, respectively, expressed as:

\begin{equation}
    \label{def_intensity_STPSP}
    \begin{aligned}
        \overline{I_{n,t}}(x,y) &= \frac{I_{n,t}(x,y)-A(x,y)}{B(x,y)} \\
        &= \cos \left( 2 \pi \left( \frac{1}{\lambda_0}-\frac{1}{T} \right) x + \phi'(x,y) \right. \\
        & \left. + \ \frac{2 t \pi}{T} + \frac{2 n \pi}{N} \right) \\
        &= \cos \left(\psi(x,y) + \frac{2 t \pi}{T} + \frac{2 n \pi}{N} \right).
    \end{aligned}
\end{equation}

\noindent From there, $\psi(x,y)$ can be inferred through a two-dimensional discrete Fourier transform, that yields

\begin{equation}
    \label{def_phi_STPSP}
    \psi(x,y) = \arg \left\{ \sum_{n=0}^{N-1} \sum_{t=0}^{T-1} \overline{I_{n,t}}(x,y) e^{-2 i \pi(\frac{n}{N} + \frac{t}{T})} \right\}.
\end{equation}

%Obtaining the phase using equation \eqref{def_phi_STPSP} is useful in order to eliminate any periodical phase error due to non-linearity, intensity saturation, and vibration (phase shifting error) \citet{2019_ri}. 

After a two-dimensional unwrapping of $\psi$, $\phi(x,y)$ is determined by adding the sampling phase to $\psi(x,y)$, in the same way as equation \eqref{def_relation_psi_phi}. Obtaining $\psi(x,y)$ using equation \eqref{def_phi_STPSP} plays a significant role in minimizing periodic phase errors arising from non-linearity, intensity saturation and vibration, as discussed by \citet{2019_ri}. This method provides greater precision than either approach (PSP or SM) individually, while preserving their respective advantages. Additionally, the ST-PSP method features a straightforward experimental setup, rapid numerical implementation, and the ability to reconstruct object height maps with high fidelity. The Python code used for all ST-PSP reconstructions in this study is available on a GitHub repository, with details provided in the $\textbf{Code availability}$ section at the end of the present study.

\subsection{The phase-to-height relation}\label{subsec2.2}

For all profilometry methods detailed in the previous section, the physically relevant information is the difference $\Delta \phi(x,y)$ between the object and the reference phases $\phi(x,y)$ and $\phi^{ref}(x,y)$, respectively, that reads

\begin{equation}
    \label{def_relation_delta_phi}
    \Delta\phi(x,y) = \phi(x,y) - \phi^{ref}(x,y).
\end{equation}

The link between $\Delta \phi$ and the object height map was first established by \citet{1983_takeda} and was then refined by \citet{2009_maurel}. The phase-to-height relation depends on whether the camera and the projector have aligned axes or not (the so-called parallel-optical or cross-optical axes geometries, respectively), and whether the projection is collimated or not. In this study, the non-collimated projection with parallel-optical-axes geometry was adopted, and the corresponding phase-to-height relationship derived by \citet{2009_maurel}, that links the surface topography height $h$ to the phase difference $\Delta \phi$, is given by:

\begin{equation}
    \label{eq_PTH_exact}
    h(x',y') = \mathscr{H}(\Delta \phi(x,y)) = \frac{L\Delta \phi(x,y)}{\Delta \phi(x,y) - \omega_0 D},
\end{equation}

\noindent where $D$ is the perpendicular distance between the projector and the camera optical axes, $L$ is the distance between the position of the entrance pupils and the reference plane, $\omega_0$ is the frequency of the projected fringes and $\mathscr{H}(X)=LX/(X-\omega_0D)$. One can also observe a shift in coordinates between both sides of equation \eqref{eq_PTH_exact}, due to parallax error. This shift is defined by

\begin{equation}
    \label{x_prime}
    x^\prime = \left( 1 - \frac{h(x^\prime,y^\prime)}{L} \right)x,
\end{equation}

\noindent and

\begin{equation}
    \label{y_prime}
    y^\prime = \left( 1 - \frac{h(x^\prime,y^\prime)}{L} \right)y.
\end{equation}

This shift is enhanced when the projector is placed closer to the reference plane or when the object is captured further away from the center of the image plane of the camera \cite{2009_maurel}. Equation \eqref{eq_PTH_exact} can be solved implicitly using regridding and interpolation of the height map from the non linear ($x^\prime$, $y^\prime$) grid to the rectangular ($x$, $y$) grid \cite{2018_grivel}. It can be noticed, however, that the experimental setup can be adapted in order for this shift to be negligible. Indeed, with a large projecting distance $L$ compared to the typical height of the object $h$, equation \eqref{eq_PTH_exact} can be well approximated by the relation 

%\begin{equation}
%    \label{eq_PTH_used}
%    h(x,y) = \frac{L\Delta \phi(x,y)}{\Delta \phi(x,y) - \omega_0 D}.
%\end{equation}

\begin{equation}
    \label{eq_PTH_used}
    h(x,y) = \mathscr{H}(\Delta \phi(x,y)).
\end{equation}

\subsection{Experimental procedure}\label{subsec2.3}

\begin{figure*}
    \centering
    \includegraphics[width=0.95\linewidth]{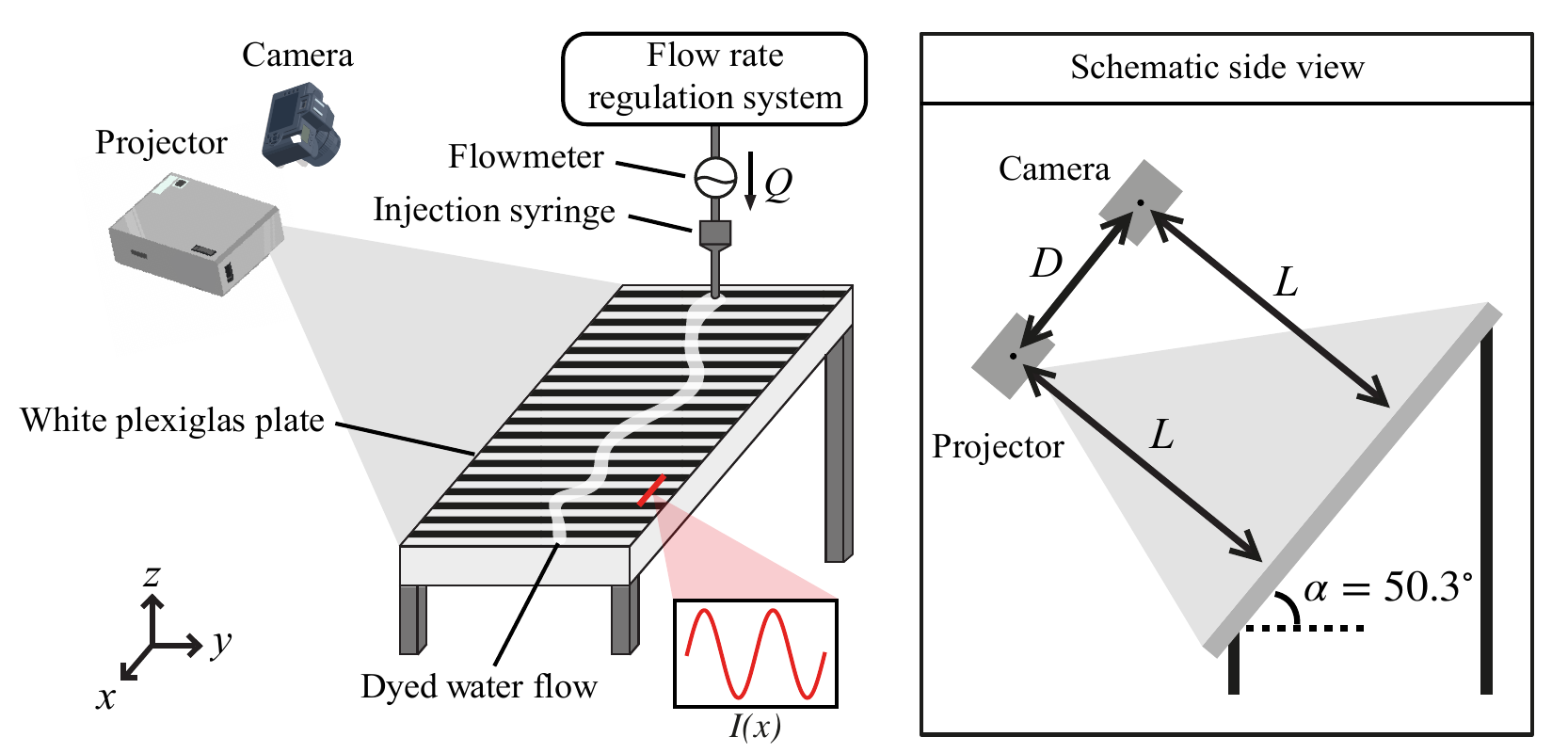}
    \caption{Schematic of the experimental setup. A thin liquid film is injected at a controlled flow rate $Q$ over a plate inclined at an angle $\alpha=50.3 ^ \circ$. The camera and projector optical axes are perpendicular to the substrate, each at a distance $L=1.32\ \rm{m}$ from it, and are separated by a distance $D=28\ \rm{cm}$. The projected fringes have an intensity $I(x)$ that follows a sinusoidal evolution along the $x$-direction.}
    \label{fig_setup}
\end{figure*}

\subsubsection{Setup}\label{subsec2.3.1}

The experimental setup of the present study is schematized in Figure \ref{fig_setup}. An Epson TW7100 video projector (4K definition) is used, alongside with a Nikon D810 camera. Depending on the dimensions of the object or flow of interest, we either used a $200~\rm{mm}$ or $50~\rm{mm}$ Nikon lens. In both cases, 36MP images are obtained. The substrate consists of a white PMMA plate (PMMA XT), chosen for its ability to provide high contrast when projecting sinusoidal fringes, thereby enhancing image projection quality. The primary criterion for selecting a substrate is to ensure that the camera can clearly capture the projected fringes. Once this condition is met, various materials — such as PMMA, metals, or wood — can, in principle, be used. One of the key advantages of the ST-PSP method is its robustness, as it remains effective even when the fringe contrast on the substrate is low or when the intensity is saturated \cite{2019_ri}. The PMMA plate thus serves as the reference plane and is positioned at an angle $\alpha=50.3^{\circ}$ to the horizontal to enable the study of different flow regimes. Both the camera and projector are positioned perpendicular to the substrate at a distance $L \simeq 1.32\ \rm{m}$ and are separated by a distance $D \simeq 28\ \rm{cm}$. To prevent any parasitical reflections on the surface of study, especially those implying significant intensity saturation, we added two crossed linear polarizers (100mm SQ TS, Edmund Optics) to the camera and projector. Furthermore, in order to minimize the time required for object reconstruction, all results of the present study were obtained using the minimal number of phase-shifted fringe patterns, \textit{i.e.}, $N=3$. Since the pictures were taken with a time interval of $\Delta t \simeq 3.3\ \rm{s}$, one reconstruction is obtained every 10 seconds, approximately. As a result, our study is limited to static or quasi-static flows within the 10-second projection time. In the case of a moving flow or object, the image projection and capturing rates can be adjusted so that the object remains quasi-static during the time required for reconstructions.

%, at a distance ranging from 1 to 5 mm

To use this optical setup for the study of water flows, two types of flow regulation systems were used depending on the flow rate $Q$. For small flow rates, $Q \in [0-50] \ \rm{mL/min}$, we use a syringe pump (Harvard Apparatus PHD 2000) with two $60 \rm{mL}$ syringes mounted in parallel. For larger flow rates, $Q \in [50-1000] \ \rm{mL/min}$, we use a recirculating hydraulic setup. A water tower is placed above the incline and is fed by a pump (Huber minichiller 600). The water castle is then connected to an injection syringe perpendicular to the substrate. A water tank is finally placed at the bottom of the substrate and re-connected to the pump in order to close the loop. An electro-valve is connected to the flowmeter to ensure the flow rate is constant. Since this setup relies on flow rate measurements with an electromagnetic flowmeter (Kobold MIM), tap water was used instead of distilled water.

\begin{figure*}
    \centering
    \includegraphics[width=\linewidth]{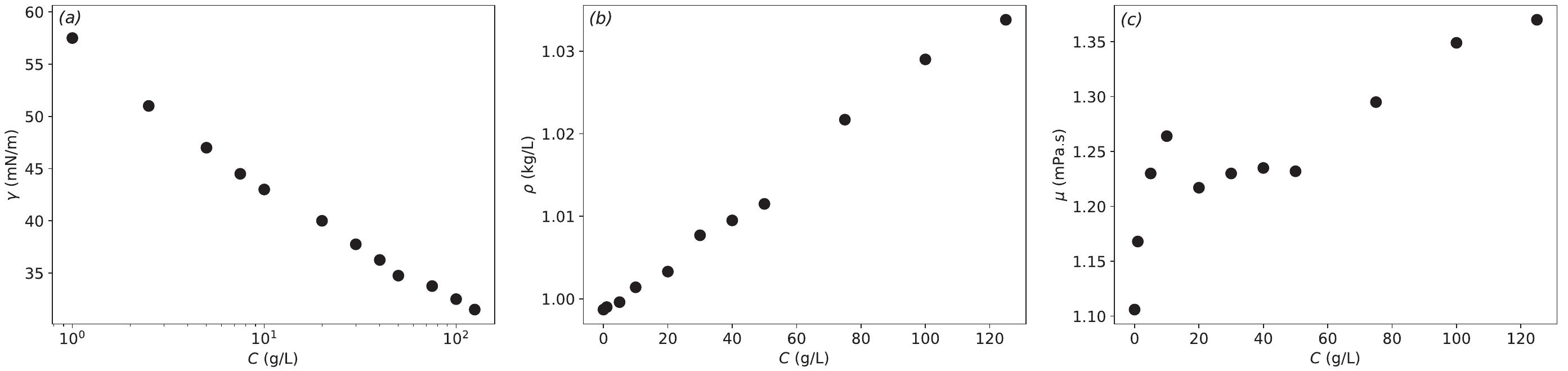}
    \caption{Evolution of the solution's hydrodynamic properties. (a) Surface tension $\gamma$, (b) density $\rho$ and (c) dynamic viscosity $\mu$ as a function of the liquid marker concentration $C$, calculated as the additional mass of marker added per liter of water.}
    \label{fig_proprietes_hydro}
\end{figure*}

In order to conduct optical profilometry based on a sinusoidal pattern, the key is to have the highest possible contrast on the object of interest. In this study, the 4K sinusoidal pattern image is encoded in shades of grey. 
Consequently, a white liquid dye, extracted from Edding 4090 markers, is added to the water solution. This dye contains water alongside titanium dioxide ($\rm{TiO_2}$) and chalk particles, with particle sizes ranging from 0.1 to 1 $\mu$m, that enhance the light diffusivity at the free surface. This additive was chosen because titania has one of the highest refractive index among natural minerals ($n=2.7$) \cite{2012_przadka}. To investigate the possible change in the hydrodynamic properties induced by the use of this marker, we measured the solution surface tension $\gamma$, density $\rho$ and dynamic viscosity $\mu$ for various liquid marker concentrations $C$ in the range $[0,\ 125] \ \rm{g/L}$, as illustrated in Figure \ref{fig_proprietes_hydro}(a)-(c). Here, the concentration $C$ is defined as the additional mass of liquid marker added per liter of water.

Significant changes in the physical properties of the water-dye solution occur as the dye concentration $C$ varies. For a given volume, adding the dye — composed of chalk (density 2.4) and titanium dioxide (density 4.2) — leads to a straightforward linear increase in the mixture’s density $\rho$, as shown in Figure \ref{fig_proprietes_hydro}(b). Additionally, Figure \ref{fig_proprietes_hydro}(a) reveals an exponential decay of surface tension $\gamma$ with $C$, with $\gamma$ decreasing from $\gamma_0 \simeq 59 \ \rm{mN.m^{-1}}$ at $C=0$ (not shown in Figure \ref{fig_proprietes_hydro}(a) due to the logarithmic concentration scale) to $\gamma_f \simeq 32\ \rm{mN/m}$ at $C = 125\ \rm{g/L}$. This trend is expected, as it is characteristic of aqueous surfactant solutions in the limit of large intermolecular distances \cite{1976_gilanyi,2021_peng}. Finally, Figure \ref{fig_proprietes_hydro}(c) displays an overall increase in viscosity $\mu$ with $C$, attributed to the rise in effective viscosity in suspensions as the volume fraction increases \cite{2018_guazzelli}.

%The surface tension of tap water (\textit{i.e.}, when $C=0$) is measured at $\gamma_0=59.5 \ \rm{mN.m^{-1}}$, and $\gamma$ undergoes an exponential decay from $\gamma_0$ to a final value $\gamma_f \simeq 32\ \rm{mN/m}$ for $C = 125\ \rm{g/L}$. This sharp decrease is most pronounced for concentrations up to 10 g/L, with a smaller decrease observed for concentrations greater than 40 g/L. In contrast, both the solution density $\rho$ and dynamic viscosity $\mu$ increase with the marker concentration, with a 4$\%$ increase in density and a 24$\%$ increase in viscosity at $C = 125\ \rm{g/L}$ compared to the case with $C = 0$. A linear dependence of $\rho$ on $C$ can also be observed.

%(in order to have sufficient electrical conductivity for the flow to be measured by the electromagnetic flowmeter)

\subsubsection{Calibration of the optical distances}\label{subsec2.3.2}

Prior to making any measurements on an object of interest, one has to determine precisely all the geometrical constants involved in the phase to height relation \eqref{eq_PTH_used}, that is $L$, $D$ and $\omega_0$. First, $\omega_0$ can be precisely and directly measured on the reference plane. However, since $L$ and $D$ can hardly be precisely determined \textit{a priori}, a numerical calibration is required. This involves using a reference object with a known geometry to determine the optimal parameters for accurate height reconstruction.

\paragraph{Principle}
To perform the calibration, a reference object with a well-defined three-dimensional geometry is used. The experimental phase difference map $\Delta \phi^{exp}$ obtained through the ST-PSP method is combined with the theoretical phase-to-height relation \eqref{eq_PTH_used} to generate a height map $h^{exp}(L,D)$. This map is then compared with precise measurements of the object's geometry (using a caliper with measurement uncertainties of $50\ \rm{\mu m}$). The optimal parameters $L_{opt}$ and $D_{opt}$ are identified by minimizing the difference between the theoretical and experimental height maps, and serve as the best values for $L$ and $D$ to accurately reconstruct an object using the phase-to-height relation \eqref{eq_PTH_used}.

%ensuring the most accurate height reconstruction of the object

\begin{figure}
    \centering
    \includegraphics[width=\linewidth]{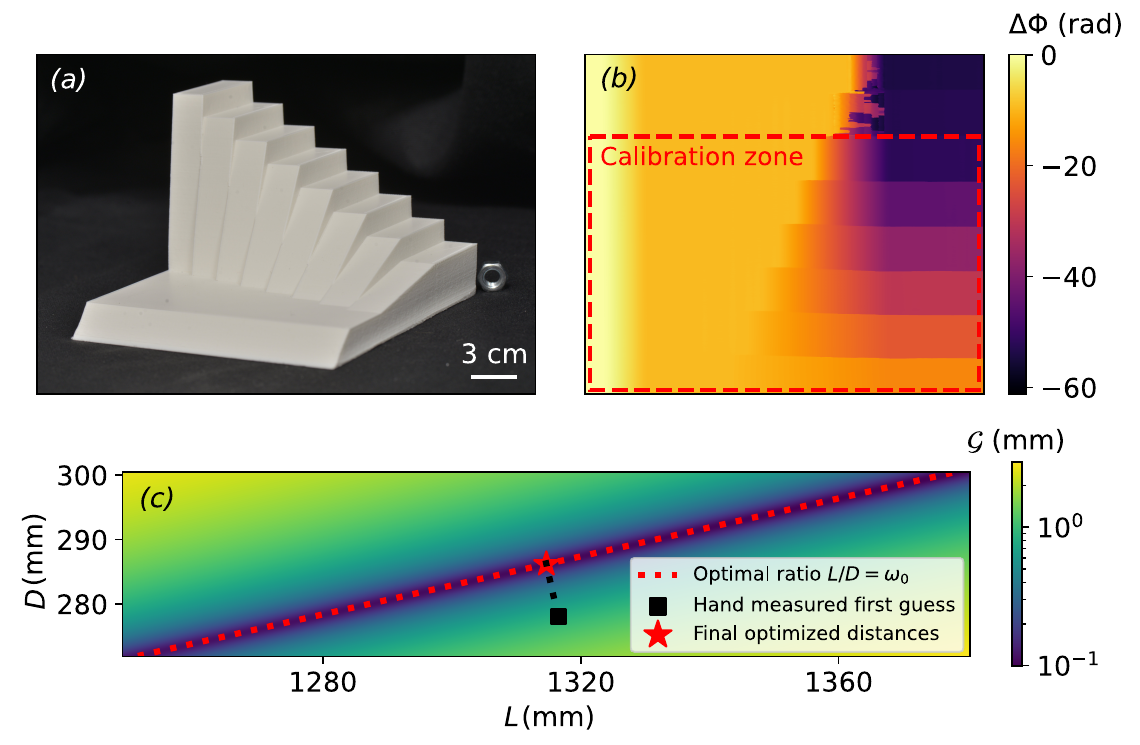}
    \caption{(a) Photograph of the calibration block. (b) Experimental phase difference $\Delta \phi^{exp}$ obtained using the ST-PSP method with a fringe pattern wavelength $\omega_0=4.15\ \rm{rad.mm^{-1}}$. (c) Minimization of the cost function ${\cal G}(L,D)$ from equation \eqref{def_cost_function} for the six slopes highlighted in red in (b) in order to find the optimized optical parameters of the setup: $L_{opt} = 1315 \ \rm{mm}$, $D_{opt} = 286 \ \rm{mm}$.}
    \label{fig_calibration}
\end{figure}

\paragraph{Calibration block design}
When calibrating a profilometry experimental setup one could choose the calibration object of their choice. However, an efficient approach is to design a wedge with a series of $K$ inclined planes with varying slopes, leading to horizontal steps of different final height. In order to prevent any discontinuities for the phase extraction, the slopes are bonded together onto a base plate which is then linearly connected to the reference substrate. In this study $K=8$ inclined planes were used and the corresponding calibration block is illustrated in Figure \ref{fig_calibration}(a). The associated three dimensional printable model is available in a dedicated repository (see the \textbf{Materials availability} section at the end of the present study for more information). 

%One simple way to do so

Horizontal steps were selected due to their simplicity: Indeed, for an object with a uniform height $h^{cst}$, the theoretical phase difference map is constant. This value, denoted $\Delta \phi^{cst}$, is derived from the phase-to-height relation \eqref{eq_PTH_used}:

%\begin{equation}
%    \label{def_methode_calibration}
%    h^{cst} = \frac{L\Delta \phi^{cst}}{\Delta \phi^{cst} - \omega_0 D}.
%\end{equation}

%\begin{equation}
%    \label{def_methode_calibration}
%    h^{cst} = \mathscr{H}(\Delta \phi^{cst}).
%\end{equation}

\begin{equation}
    \label{def_methode_calibration_1}
    \Delta \phi^{cst} = \omega_0D \frac{h^{cst}}{h^{cst}-L}.
\end{equation}

Consequently, one of the simplest and most efficient methods to calibrate the optical setup is to measure the average phase difference value for each plateau, input it into the theoretical phase-to-height relation \eqref{eq_PTH_used}, and minimize the absolute difference of the resulting height with the block's caliper measurements $h^{mes}$. To improve the accuracy of the comparison, the $i$-th plateau ($i \in [1,K]$) is divided into $N_i$ segments, where $N_i$ represents the number of pixels corresponding to the width of the plateau. For plateau $i$ and segment $j$ ($j \in [1,N_i]$), the experimental phase difference $\Delta \phi^{exp}$ is averaged over the plateau length, resulting in the value $\Delta \phi^{exp}_{i,j}$. This phase difference is then converted into a theoretical height $h^{th}_{i,j}$ using \eqref{eq_PTH_used}, enabling a comparison with the measured height $h^{mes}_i$ of the $i$-th plateau. For the $j$-th segment of the $i$-th slope, the theoretical height $h^{th}_{i,j}$ is expressed as

%\begin{equation}
%    \label{def_methode_calibration}
%    h^{th}_{i,j}(L,D) = \frac{L\Delta \phi^{exp}_{i,j}}{\Delta \phi^{exp}_{i,j} - \omega_0 D}.
%\end{equation}

\begin{equation}
    \label{def_methode_calibration_2}
    h^{th}_{i,j}(L,D) = \mathscr{H}(\Delta \phi^{exp}_{i,j}).
\end{equation}

\noindent Using these definitions, the final optimized parameters $L_{opt}$ and $D_{opt}$ are determined by minimizing the cost function

\begin{equation}
    \label{def_cost_function}
    {\cal G} (L,D) = \sqrt{\frac{1}{K} \sum_{i=1}^{K} \frac{1}{N_i} \sum_{j=1}^{N_i} \left( h^{mes}_{i} -  h^{th}_{i,j}(L,D) \right)^2}.
\end{equation}

%A key consideration is

\paragraph{Slope limitation}
It should be mentioned that not all surface slopes can be accurately reconstructed using fringe projection profilometry methods. One fundamental limitation arises from the nonlinear response of both the camera and the projector, which distorts the fringe profiles due to the nonlinear mapping of the projector input to the captured image intensity. This distortion introduces higher-order harmonics of the projected fringe frequency $f_0$, constraining the available frequency range for phase gradients and ultimately imposing an upper bound on the measurable surface slope. For the FTP method with parallel optical axes and non-collimated light, for instance, this limit is

\begin{equation}
    \label{limit_gradient_FTP}
    \left| \frac{\partial h}{\partial x} \right|_{FTP} < \frac{L}{3D},
\end{equation}

\noindent a condition that ensures separation of the fundamental frequency component from both background variations and higher-order harmonics \cite{1983_takeda}. However, when the second harmonic is absent — which is achievable, for example, through quasi-sinusoidal projection and $\pi$-phase shifting — \citet{2001_Su} demonstrated that the fundamental component can extend to $2f_0$ without overlap, leading to a less restrictive limit of $L/D$.

%as employed in this study
%as in the present study

In the experiments conducted in this study, the second harmonic was not observed, suggesting that the upper bound on the measurable slope may follow a similar constraint to the one identified by \citet{2001_Su} for $\pi$-phase shifting under equivalent experimental conditions (\textit{i.e.}, parallel optical axes, non-collimated projection, and negligible parallax errors). Through experimental investigation, it was found that the slope limit for the ST-PSP method can be approximated by

\begin{equation}
    \label{limit_gradient}
    \left| \frac{\partial h}{\partial x} \right| < \xi \frac{L}{D},
\end{equation}

\noindent with $\xi \simeq 0.7 \pm 0.05$. As a result, the calibration block's slopes $\alpha_i$ should not exceed the critical value defined by $\alpha_{cr}= \arctan \left(\xi L/D \right)$.

\paragraph{Results}

For our specific setup, the calibration block is shown on Figure \ref{fig_calibration}(a) and consists of a series of eight inclined planes with slopes ranging from $11.3^\circ$ to $80^\circ$, leading to horizontal plateaus with final heights ranging from 1.75 cm to 7 cm. The projected fringe frequency is $\omega_0=4.15\ \rm{rad.mm^{-1}}$ and the experimental phase difference $\Delta \phi^{exp}$ obtained with the ST-PSP is shown in Figure \ref{fig_calibration}(b). The shape of the height map closely resembles that of the physical block, as it is divided into sections with varying slopes, all continuously connected to the substrate on the left side, where $\Delta \phi^{exp}=0$. It can be observed that the two steepest slopes located at the top of the block ($75^\circ$ and $80.5^\circ$, respectively) are poorly defined and affected by noise. This highlights the slope limitations mentioned in the previous paragraph as the critical reconstructible slope in this case will turn out to be $\alpha_{cr} = 72.7^\circ$. Therefore, only the six smallest slopes of the block, ranging from $11.3^\circ$ to $68.2^\circ$, are considered thereafter, as they satisfy the sampling condition \eqref{limit_gradient}. According to the previous definitions, this corresponds to $K=6$. The area encompassing these slopes is highlighted by the red dashed line in Figure \ref{fig_calibration}(b).

To minimize the cost function $\cal G$, initial estimates for both $L$ and $D$ were obtained through manual measurements, yielding $L_{\text{init}} = (1320 \pm 40) \ \rm{mm}$ and $D_{init}=(280 \pm 20) \ \rm{mm}$. Given the uncertainties of those measurements, which do not exceed $8\%$, the cost function $\cal G$ is minimized within a range of $\pm10\%$ relative to the first guesses. The corresponding two-dimensional cost function map for the calibration test is shown in Figure \ref{fig_calibration}(c). A continuum of local minima can be observed, as indicated by the red dashed line, corresponding to a constant ratio $L/D = \omega_0$. This red dashed line corresponds to a linearized phase-to-height relation in the case $\Delta \phi \ll \omega_0 D$. The final optimized values $[L_{opt}, D_{opt}]$ are determined by the orthogonal projection of $[L_{init}, D_{init}]$ onto the red dashed line, ensuring they lie within the initial uncertainty range of the first guess measurements. For the calibration test, this resulted in $L_{opt} = 1315 \ \rm{mm}$ and $D_{opt} = 286 \ \rm{mm}$, yielding an average error of $0.1\ \rm{mm}$ across all segments.

\subsubsection{Optimizing the optical setup for the object of study}\label{subsec2.3.3}

Investigating millimetric heights, such as those typically observed in capillary flows, requires adjusting the experimental setup to match the dimensions of the studied object. This involves optimizing three key constants: $L$, $D$, and the projected fringe wavelength $\lambda_0$.

\paragraph{Optimizing $L$ and $D$}

The first step is to select a camera lens with sufficient zoom to accurately detect intensity variations on the object of interest. Next, the distance $D$ should be adjusted to allow observation of fringe displacement at the lower end of the object’s height distribution. However, increasing $D$ reduces the maximum reconstructible height gradient, as described by equation \eqref{limit_gradient}. Therefore, $D$ should be adapted based on which of these conditions is most critical. Additionally, when studying small-scale objects such as capillary flows, the projection distance $L$ (along with its throw angle) should be minimized to maximize the number of projected pixels per unit length. This increases contrast on the projection surface and minimizes projection non-linearity, ensuring that the projected pattern remains as close as possible to a perfect sine wave on a flat surface. For very short projection distances and taller objects, parallax distortions may occur, requiring the use of the exact phase-to-height relation \eqref{eq_PTH_exact} to accurately extract the height map.

\begin{figure*}
    \centering
    \includegraphics[width=0.95\linewidth]{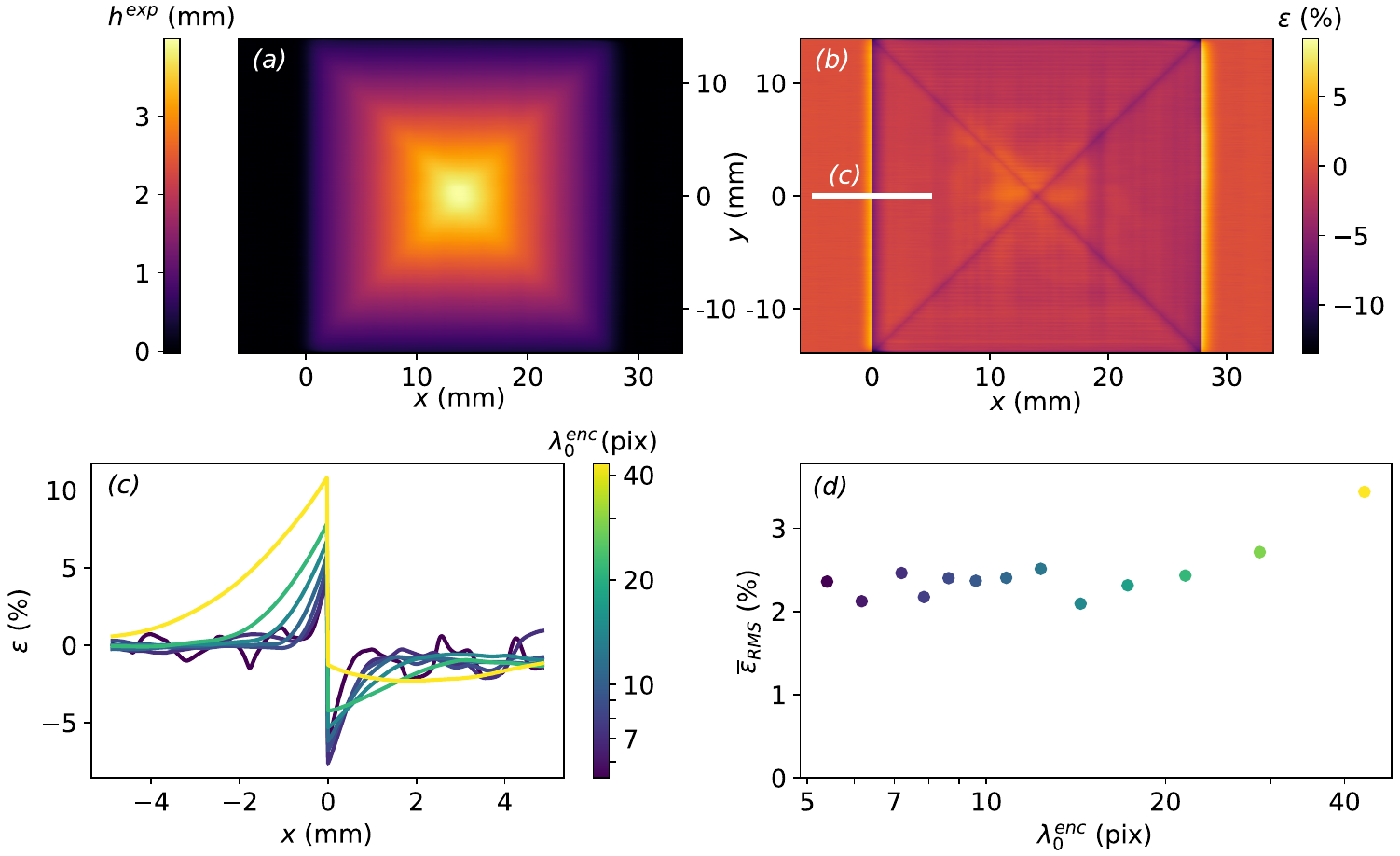}
    \caption{Influence of the fringe pattern wavelength $\lambda_0$ on the reconstruction of a 3D-printed white PMMA pyramid ($28 \times 28 \times 4.1\ \rm{mm^3}$). (a) ST-PSP reconstruction using $\lambda_0^{enc} = 10.8 \ \rm{px}$ for the projected image ($\lambda_0=1.51\ \rm{mm}$ when projected on the substrate), and (b) the corresponding error map. The white line highlighted in the error map (b) is plotted for various encoded wavelengths in (c), illustrating how the ST-PSP method handles physical height discontinuities at the pyramid edges. (d) Averaged normalized error $\overline{\varepsilon}_{RMS}$ on the pyramid as a function of $\lambda_0$. The color of the symbols corresponds to the colorbar in (c).}
    \label{fig_etude_lambda}
\end{figure*}

\paragraph{Optimizing the pattern wavelength $\lambda_0$}

Once $L$ and $D$ are appropriately determined through the procedure explained in section \ref{subsec2.3.2}, the next step is to adjust the wavelength of the projected fringe pattern. To determine the optimal wavelength for the present study, a sensitivity analysis was conducted using a 3D-printed pyramidal object. Since this study focuses on capillary flows, with typical heights between 0.4 and $4\ \rm{mm}$, the pyramid was designed with a comparable height field, reaching $4.1\ \rm{mm}$ at its peak. With $L$ and $D$ held constant, varying the projected wavelength $\lambda_0$ (in mm) directly translates to changes in the encoded wavelength of the projected image, $\lambda_0^{enc}$ (in pixels). Therefore, all results are presented as a function of $\lambda_0^{enc}$ instead of $\lambda_0$.

A typical reconstruction of the pyramid using the ST-PSP method is shown in Figure \ref{fig_etude_lambda}(a). This reconstruction corresponds to a case where $\lambda_0^{enc}~=~10.8~\rm{px}$ on the encoded 4K image, which corresponds to $\lambda_0=1.51~ \rm{mm}$ on the reference plane. The results are represented by the experimental height map $h^{exp}$. To assess the accuracy of this reconstruction relative to the actual height profile of the object, the normalized error $\varepsilon$ is defined as follows:

\begin{equation}
    \label{eq_err_pyr}
    \varepsilon(x,y) = \frac{h^{exp}(x,y)-h^{th}(x,y)}{h^{th}_{max}},
\end{equation}

\noindent with $h^{th}$ the physical height distribution of the pyramid, and $h^{th}_{max}$ its maximal value. The error map obtained with $\lambda_0^{enc}=10.8\ \rm{px}$ is displayed in Figure \ref{fig_etude_lambda}(b). For this example, $\varepsilon$ almost never exceeds the $\pm 5\%$ mark since the main mismatches happen on the edges of the pyramid and on the sharp gradients separating each face. This is due to the interpolation performed by the SM algorithm inside the ST-PSP method, that tends to smoothen gradient discontinuities. Since the interpolation is performed along the $x$-axis, this smoothing error is the highest for large gradient jumps in this direction, such as the left and right edges of the pyramid. In order to highlight this effect, a white line is drawn in Figure \ref{fig_etude_lambda}(b) which crosses the pyramid's left edge discontinuity in $x=0$. Figure \ref{fig_etude_lambda}(c) shows the error along this line for various projected image wavelengths, ranging from $5.4\ \rm{px}$ (blue) to $43.2\ \rm{px}$ (yellow), corresponding to $\lambda_0$ values between 0.75 and $6.05\ \rm{mm}$. The divergence in $x=0$ is due to the physical discontinuity of $h^{th}$ along the edge of the pyramid. One can observe that increasing the encoded fringe wavelength $\lambda_0^{enc}$ results in both a higher and wider divergence in $x=0$. However, $\lambda_0^{enc}$ is constrained by the physical limitations of the projector. Given the resolution of the projector (4K for the Epson TW7100), decreasing $\lambda_0^{enc}$ can lead to significant degradation in the quality of the encoded periodic pattern. In this sensitivity analysis, the limitation is encountered near the lower bound of the study, around $\lambda_0^{enc} \simeq 5\ \rm{px}$. As illustrated in Figure \ref{fig_etude_lambda}(c), an irregular error pattern appears at this wavelength, and further reducing $\lambda_0^{enc}$ would result in increased instability and a sharp rise in absolute error $\left| \varepsilon \right|$.

In order to analyze the overall performance of the ST-PSP method for different encoded wavelengths, one can define an averaged squared quadratic error $\overline{\varepsilon}_{RMS}$ over the entire surface of the pyramid $S$ as

\begin{equation}
    \label{eq_err_tot_pyr}
    \overline{\varepsilon}_{RMS}=\sqrt{\frac{1}{S}\iint_{S} \varepsilon^2\, \mathrm{d}S}.
\end{equation}

\noindent The evolution of $\overline{\varepsilon}_{RMS}$ with $\lambda_0^{enc}$ is shown in Figure \ref{fig_etude_lambda}(d). Notably, this averaged error never exceeds 4$\%$, even for the largest encoded wavelengths of study where smoothing effects are most pronounced. This demonstrates the robustness of the method in reconstructing an object's surface, regardless of the encoded fringe wavelength. Surprisingly, $\overline{\varepsilon}_{RMS}$ remains constant for $\lambda_0^{enc} \in [5, 20]\ \rm{px}$, despite the increase in smoothing error with $\lambda_0$. However, to achieve the highest fidelity in reconstructing areas with high gradients, $\lambda_0$ should be minimized as much as possible within the physical limitations of the projector. In this study, based on the projector specifications, this corresponds to a fringe wavelength $\lambda_0^{enc} \simeq 6-7 \ \rm{px}$ on the encoded 4K image.

\subsection{Optimizing the white dye concentration}\label{subsec2.4}

\begin{figure}
    \centering
    \includegraphics[width=\linewidth]{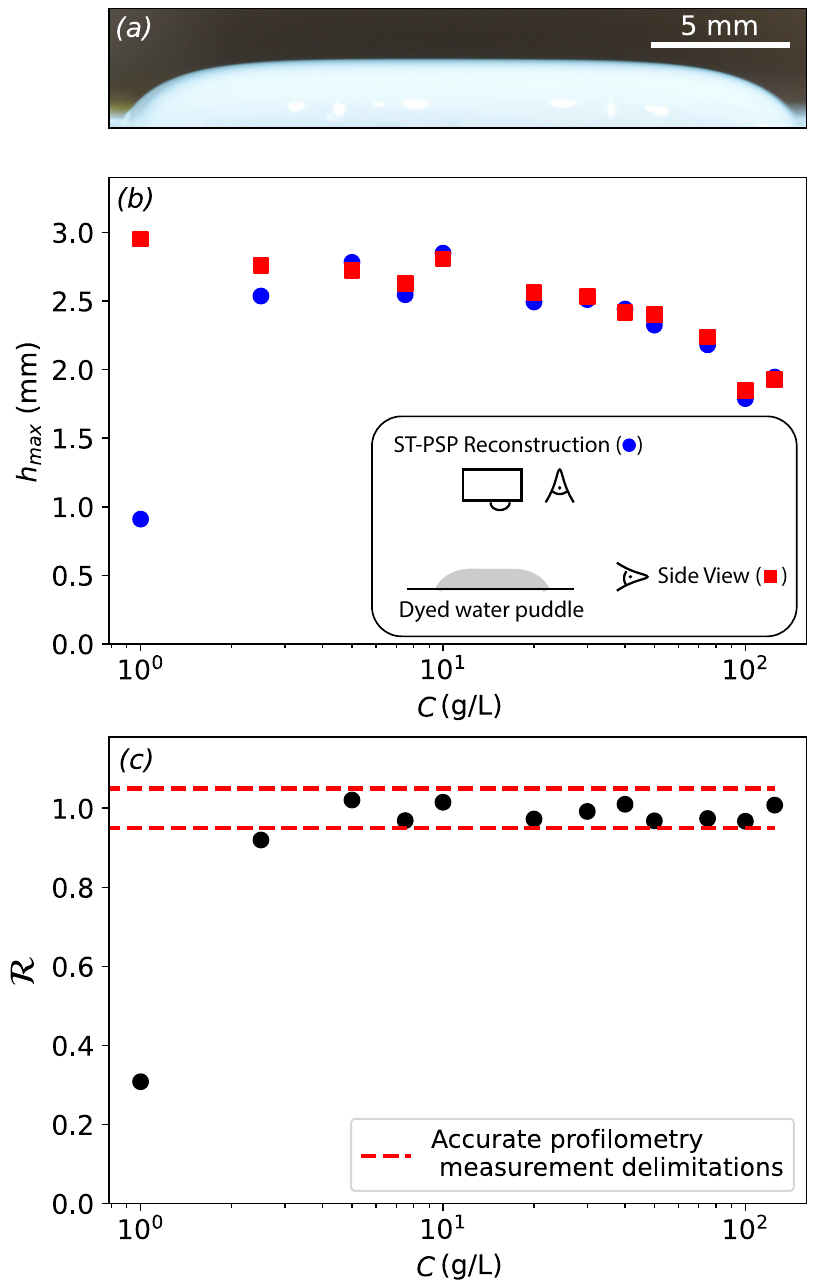}
    \caption{Quantitative analysis of the minimum concentration of white marker required to ensure sufficient light diffusivity of the free surface. (a) Side view example of a puddle with $C=10 \ \rm{g/L}$. (b) Maximum height, $h_{max}$, estimated either using the ST-PSP method (\textcolor{blue}{\ding{108}}) or a direct measurement with a side view camera (\textcolor{red}{\ding{110}}), as a function of the marker concentration $C$. (c) Evolution of the ratio $\cal R$ between the maximum height captured by the ST-PSP, $h^{STPSP}_{max}$, and the one obtained using the side view camera, $h^{SV}_{max}$. The free surface is considered diffusive enough enough once $\cal R$ lies within the interval $[0.95,1.05]$.}
    \label{fig_palliers_concentration}
\end{figure}

Once the method is properly calibrated on a block with optimal parameters $L$, $D$ and $\lambda_0$, it can be implemented in order to study the free surface of flows. The final parameter to optimize is the concentration $C$ of white dye, which must be high enough to ensure that the free surface of the flow is sufficiently diffusive. One way to quantitatively verify that the concentration is adequate is to compare the reconstructions of a known liquid geometry for different concentrations of white marker. In the following study, these tests have been performed on a still water puddle of typical dimensions $2.7 \ \rm{cm}\times2\ \rm{cm}$. Hydrophobic tape was used in order to prevent the puddle from spreading over time. The results are presented in Figure \ref{fig_palliers_concentration}, where the maximum height of the puddle was determined thanks to the ST-PSP reconstructions (blue disks) as well as with a side view camera (red squares).

A discrepancy appears between the ST-PSP and the side-view camera measurements when $C < 4\ \rm{g/L}$. This difference is expected because of the intrinsic absorption and scattering of light within the suspension. This phenomenon can be described through an extinction coefficient $\zeta$, defined as the inverse of the mean penetration depth of incoming light beams. In a titanium dioxide suspension, scattering is the dominant contributor to this extinction coefficient \cite{1996_cabrera}. According to Mie scattering theory, for a monodisperse suspension in an unsaturated medium, the extinction coefficient $\zeta$ evolves linearly with the particle concentration $C$ \cite{1984_baker, 2024_Francisco}. Indeed, \citet{2024_Francisco} evidenced that

\begin{equation}
    \label{light_attenuation_coefficient}
    \zeta \propto \dfrac{1}{l^*} \propto C,
\end{equation}   

\noindent with $l^*$ the transport mean free path of the suspension, which is inversely proportional to the particle concentration $C$. Consequently, the average penetration distance of light rays increases as the particle concentration decreases, becoming non-negligible at very low concentrations when compared to millimetric heights, such as the height of the puddle shown in Figure \ref{fig_palliers_concentration}(a). This explains why the reconstructed heights at low concentrations are lower than those measured by the side view camera and why this discrepancy diminishes as the dye concentration increases.

One can also notice in Figure \ref{fig_palliers_concentration}(a) that the maximum height $h_{max}$ detected by the side view camera decreases with $C$. Indeed, as seen in Figure \ref{fig_proprietes_hydro}(a), the surface tension of the solution decreases sharply as the marker concentration increases. As a result, when $C$ increases the capillary length decreases, so does the maximum height. The reconstructed height map is considered accurate if the ratio ${\cal R} = h^{STPSP}_{max}/ h^{SV}_{max}$ between the maximum height $h^{STPSP}_{max}$ detected with the ST-PSP and the one measured using the side-view camera, $h^{SV}_{max}$, satisfies ${\cal R} \in [0.95,1.05]$. This ratio is shown in Figure \ref{fig_palliers_concentration}(c), which indicates that accurate ST-PSP reconstructions are achieved when the marker concentration exceeds $C_{lim} \simeq\ 5\ \rm{g/L}$. Note that, according to the previous discussion, the concentration threshold can vary depending on the liquid object. Therefore, prior to studying a specific flow, we recommend calibrating the concentration of the white marker with a liquid geometry similar to that of the flow of interest.

%As $C$ increases, the puddle geometry changes, with a reduction in contact angle and consequently a decrease in maximum height
 
\section{Application to capillary flows}\label{sec3}

%In this section, different thin liquid films forming a contact line with their substrate are studied using the ST-PSP method.
%These flows usually feature a complex interplay between inertia, capillarity, wettability and viscous effects \cite{2005_mertens,2006b_le_grand-piteira,2016_josserand}.

\begin{figure}
    \centering
    \includegraphics[width=\linewidth]{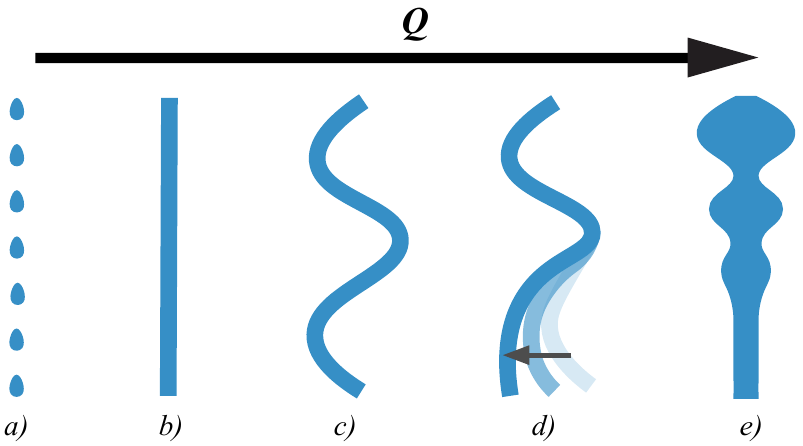}
    \caption{Illustration of different capillary flows over an incline. In increasing order of the flow rate: (a) Dripping regime, (b) straight rivulet, (c) stationary meander, (d) dynamic meander and (e) braided film.}
    \label{fig_types_ecoulements}
\end{figure}

When fluid is injected at a certain flow rate $Q$ on an incline, different regimes of capillary flows can be observed such as drops, rivulets, steady or dynamic meanders and braided liquid films \cite{1984_nakagawa,1992_nakagawa,2006b_le_grand-piteira,2022_monier}, as illustrated in Figure \ref{fig_types_ecoulements}. Previous studies have determined the shapes of moving or impacting droplets \cite{2015_hu,2012_lagubeau} as well as straight rivulets \cite{2013_isoz}. However, until now, there has been no accurate method for reconstructing the three-dimensional shape and apparent contact angles of capillary flows forming a contact line with their substrate. In this section, the previously characterized ST-PSP method and the experimental setup illustrated in Figure \ref{fig_setup} are employed, except in \ref{subsec3.2} where the setup was tilted to study a horizontal sessile drop. Depending on the typical dimensions of the flow, we used different camera lenses in order to maximize the precision of the reconstructions.

\subsection{Optimization of the liquid dye concentration}\label{subsec3.1}

\begin{figure}
    \centering
    \includegraphics[width=\linewidth]{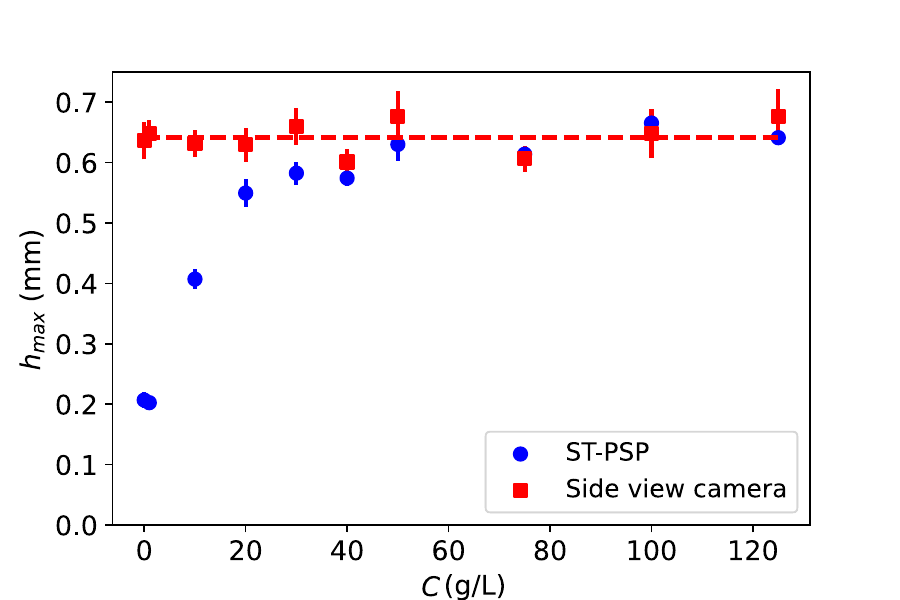}
    \caption{Quantitative analysis of the minimum concentration of white marker required to ensure sufficient light diffusivity for the free surface of a rivulet flowing at a rate $40\ \rm{mL/min}$. The maximum reconstructed height matches the maximum height seen by a side view camera for concentrations $C \geqslant C_{lim} \simeq 40\ \rm{g/L}$.}
    \label{fig_concentration_ruisselet}
\end{figure}

As outlined in the previous section, determining the minimal dye concentration is essential for profiling a flow free surface. Since this concentration threshold depends on the size of the studied object, a calibration was performed using a reference object with similar characteristics to the capillary flows of interest. This calibration involved a straight rivulet flowing over an inclined plane at a flow rate of  $Q=40\ \rm{mL/min}$, using the experimental setup depicted in Figure \ref{fig_setup}. Similar to the approach in Figure \ref{fig_palliers_concentration}, a side-view camera was used to compare the results with the ST-PSP reconstructions. In Figure \ref{fig_concentration_ruisselet}, the maximum heights from both the ST-PSP and the side-view camera are presented as a function of the concentration $C$. For $C \geqslant C_{lim} \simeq 40\ \rm{g/L}$, the reconstructed and observed heights align within measurement uncertainty intervals and the criteria ${\cal R} = 1\pm0.05$ defined in paragraph \ref{subsec2.4} is verified. This threshold is higher than the previous puddle threshold ($C_{lim} \simeq 5\ \rm{g/L}$) because the rivulet height (approximately $0.7 \ \rm{mm}$) is significantly lower than the puddle height (approximately  $2.5 \ \rm{mm}$). As a result, the scattering extinction length must be much smaller in this case to remain negligible compared to the rivulet height, leading to a much higher threshold concentration following equation \eqref{light_attenuation_coefficient}. Based on these findings, all subsequent experiments were conducted using a concentration of $C^{exp}=50\ \rm{g/L}$.

%, thus justifying the $50\ \rm{g/L}$ concentration employed in this study

\subsection{Sessile drop}\label{subsec3.2}

\begin{figure*}
    \centering
    \includegraphics[width=0.95\linewidth]{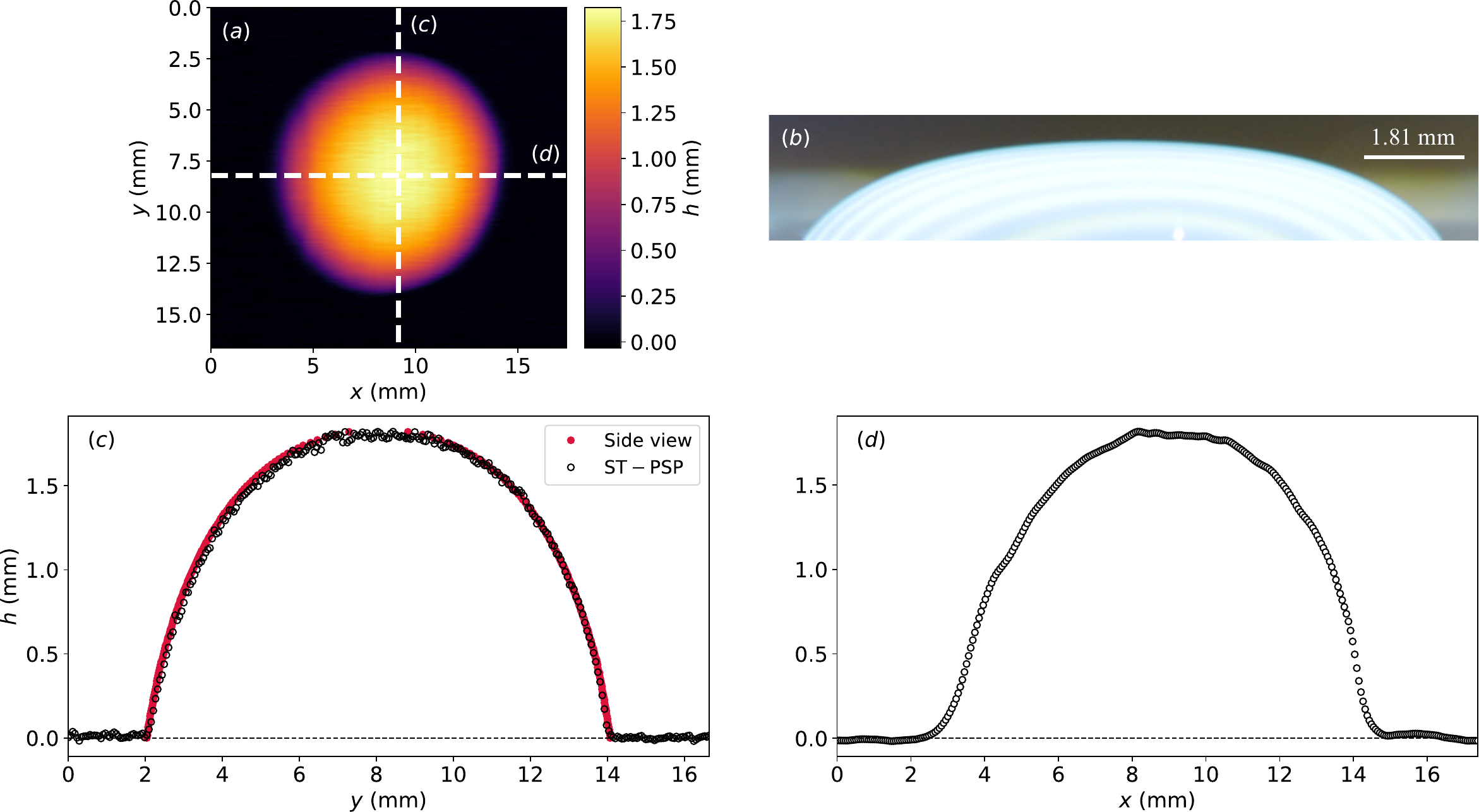}
    \caption{(a) Height distribution of a sessile droplet obtained using the ST-PSP method. The local elevation (in millimeters) is indicated by the color bar. (b) Side view photograph of the same droplet. The thick white bar sets the scale. (c) Sections in the $\left( y, z \right)$ plane for $x \simeq 9.17$ mm. The black and red circles correspond to the free surface extracted from (a) and (b), respectively. (d) Section from the reconstruction in (a), in the $\left( x, z \right)$ plane with $y \simeq 8.24$ mm. In (c) and (d), the black dashed line corresponds to $z=0$.}
    \label{fig_sessile_drop}
\end{figure*}

The first situation reported here corresponds to a sessile drop that has been deposited on a horizontal PMMA surface ($\alpha=0^{\circ}$). As the diameter of the studied drop ($d \simeq 10 \ \rm{mm}$) is considerably smaller than the largest dimensions of subsequent capillary flows, the experimental setup was adjusted as outlined in \ref{subsec2.3.3}. The camera and projector were brought closer to the substrate to ensure that the pattern wavelength remains negligible compared to the drop diameter. The calibration of the setup resulted in this case in $L_{opt} = 650 \ \rm{mm}$ and $D_{opt} = 278 \ \rm{mm}$, with a pattern wavelength on the reference substrate of $\lambda_0 = 0.91 \ \rm{mm}$.

After deposition, the drop can be considered quasi-static, since its spreading time is much longer than the time needed for both the ST-PSP reconstruction and direct acquisitions \cite{1992_hocking}. The ST-PSP method, as described in section \ref{sec2}, is applied on the resting droplet and the outcome obtained in terms of the height distribution is illustrated in Figure \ref{fig_sessile_drop}(a), with the colorbar indicating the local height measured in millimeters. The area corresponding to the reference plane always has an elevation very close to zero, as highlighted by the darker shades in Figure \ref{fig_sessile_drop}(a). In the quasi-circular wetted area, a clear signal corresponding to the presence of the liquid is detected by the ST-PSP processing with a maximal height $h_{max} \simeq 1.8\ \rm{mm}$. Here, $h_{max}$ is well above the typical noise level measured at the reference plane, that is about $20\ \rm{\mu m}$. This value is  reported in Table \ref{table_1} alongside the corresponding ratio between the maximum height and the noise level, $\rm{SNR_{max}} = 91$. In order to have a reference image for comparison, a side view image of the sessile drop is also recorded, as illustrated in Figure \ref{fig_sessile_drop}(b). Figure \ref{fig_sessile_drop}(c) shows the section of the extracted ST-PSP height map (a) for which $x=9.17\ \rm{mm}$, represented by black circles, while Figure \ref{fig_sessile_drop}(d) presents the section where $y=8.24\ \rm{mm}$.

The side-viewed free surface position extracted from Figure \ref{fig_sessile_drop}(b) is also represented by red circles in Figure \ref{fig_sessile_drop}(c). The two profiles show a good agreement, demonstrating that the spherical cap shape of the drop is accurately captured by the profilometry method. The free surface captured by the ST-PSP method features a large value for $\rm{SNR_{max}}$, which results in very small fluctuations in the height profile obtained. The agreement between the two curves highlights the relevance of the ST-PSP method for the fine measurement of such a capillary object, and allows us to determine its relevant characteristics such as its maximal height, its apparent contact angles with the substrate and its local curvature. In particular, the left and right contact angles that the liquid forms with the substrate have been extracted for both profiles, using the methodology developed by \citet{2020_quetzeri-santiago}: a quadratic polynomial is used to fit the contact line for the reference free surface (red circles), leading to left and right contact angles of $(57.0 \pm 1.1)^\circ$ and $(58.6 \pm 1.0)^\circ$, respectively. On the other hand, we applied a small Savitzky-Golay filter (10-pixel window with a quadratic polynomial) on the ST-PSP profile to reduce as much as possible the noise level without modifying significantly the signal information. Then, using a third-order polynomial to fit the contact line, the left and right contact angles are found to be $(53.3 \pm 1.4)^\circ$ and $(54.8 \pm 1.1)^\circ$, respectively. Given the large sensitivity of the method from \citet{2020_quetzeri-santiago} with respect to the accurate detection of the triple point, the relatively good agreement between the two contact angle estimates indicates that the ST-PSP is efficient for measuring contact angles, which is of great interest for a wide range of problems involving capillary flows.

\begin{figure*}
    \centering
    \includegraphics[width=0.8\linewidth]{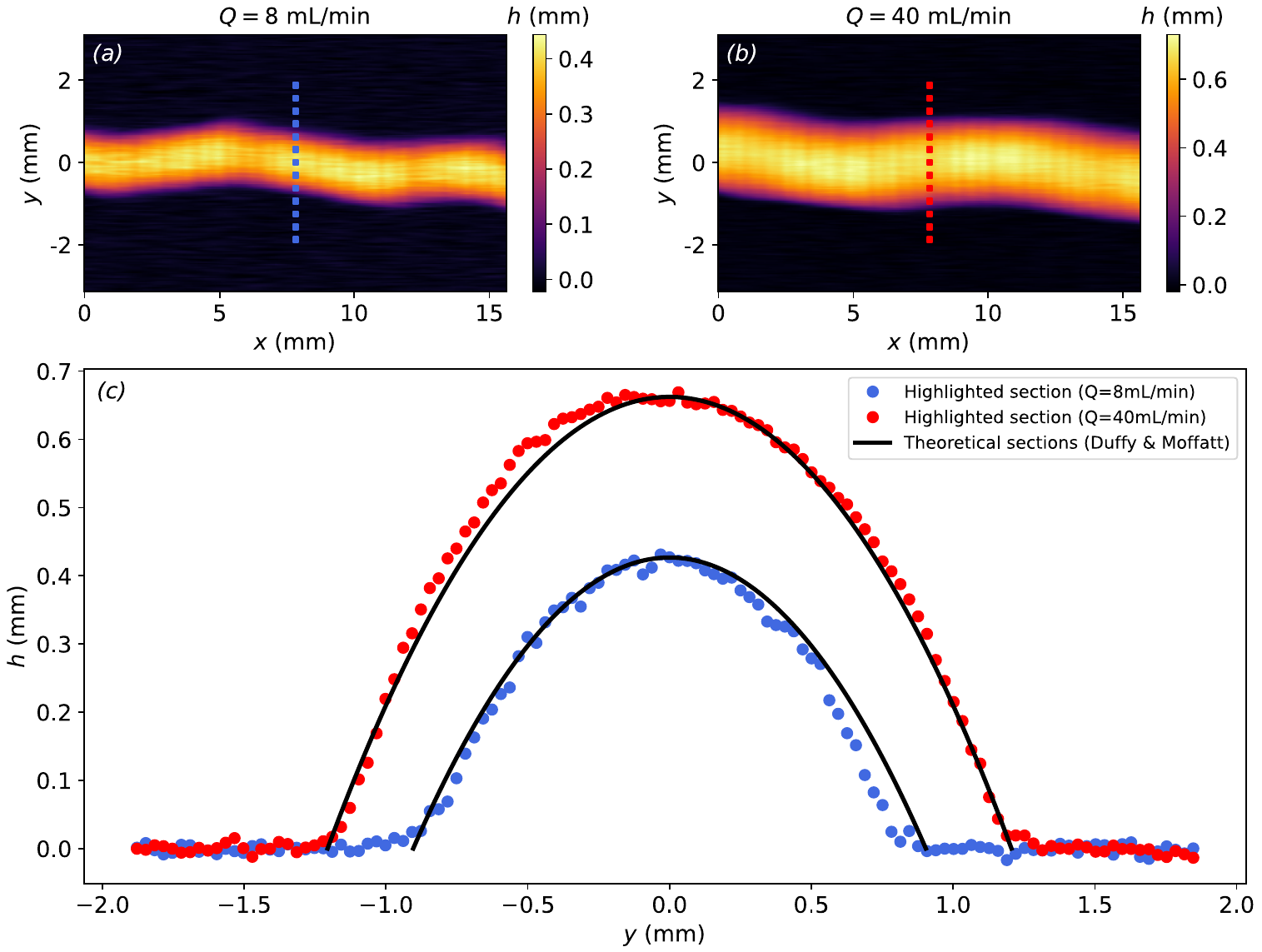}
    \caption{Three dimensional height maps obtained using the ST-PSP method for scanning straight rivulets flowing at a rate of (a) $Q \simeq 8\ \rm{mL/min}$ and (b) $Q \simeq 40\ \rm{mL/min}$ over a plate inclined at $50.3^\circ$ with the horizontal. The colorbar shows the local elevation, and the rivulet flows from left to right. The blue and red dashed lines represent the experimental sections plotted in (c), with the blue and red circles corresponding to the experiments reported in (a) and (b), respectively. The black solid lines are the analytical predictions from \citet{1995_duffy}, obtained using equations \eqref{dm_eq_1}-\eqref{dm_eq_2} and the maximum height of the experimental rivulet.}
    \label{fig_rivulet}
\end{figure*}

It should be noted that the ST-PSP method is anisotropic, because the SM component of the algorithm involves down-sampling and interpolation along the $x$-direction, which effectively applies a one-dimensional lowpass filter to the data. This results in reduced noise when plotting profiles in the $x$-direction, creating a smoothing effect that depends on the sampling period. The smoothing effect worsens with larger fringe wavelengths $\lambda_0$, leading to a greater loss of precision when analyzing contact lines or sharp height gradients in the direction of the fringes. This is demonstrated in Figure \ref{fig_sessile_drop}(d), where the contact line appears smoother compared to Figure \ref{fig_sessile_drop}(c). However, it can be noticed that the maximum height remains largely unaffected in this case because the fringe pattern wavelength remains sufficiently small ($\lambda_0 = 13\ \rm{px} \simeq 0.3\ \rm{mm}$). To ensure high precision for sharp gradient measurements, the fringes should be oriented perpendicular to the direction of interest. This makes the ST-PSP method particularly well-suited for studying capillary flows aligned with the direction of the projected fringes, such as rivulets. If the surface gradients are more significant in a different direction, the orientation of the optical setup can be adjusted to align them with the direction of the projected fringes.

%It should be noted that, for the profile in (c), there is no spatial smoothing of the data due to the profilometry method, unlike in figure \ref{fig_sessile_drop}(d) that corresponds to a section in the $\left( x, z \right)$ plane with $y=8.24\ \rm{mm}$ (\textit{i.e.}, taken close to the center of the droplet). Indeed, in that case, the Sampling Moiré part of the ST-PSP algorithm operates a spatial averaging which appears to act as a bandpass filter on the data, as illustrated by the absence of noise in the signal. Nevertheless, the maximum detected height is the same as in figure \ref{fig_sessile_drop}(c), and it is found that the main effect of this averaging is to smoothen the height distribution close to the contact line in the $x$-direction. As a result, if apparent angle measurements are relevant in planes perpendicular to the direction of the projected fringes, this is not the case for parallel planes. This means the ST-PSP method is particularly well-suited to study capillary objects that flows in the direction of the projected fringes such as, for instance, rivulets.

\subsection{Straight rivulets}\label{subsec3.3}

For the experiments that follow, the PMMA plate is inclined at an angle of $\alpha = 50.3^\circ$, as illustrated in Figure \ref{fig_setup}. The camera and projector are placed at a distance $L_{opt} = 1315 \ \rm{mm}$ from the substrate, while the distance between the two devices is $D_{opt} = 286 \ \rm{mm}$, as established in the calibration section \ref{subsec2.3.2}.

Firstly, we consider the case of rivulets flowing over the incline at two representative flow rates of $Q = 8\ \rm{mL/min}$ and $Q = 40\ \rm{mL/min}$. The corresponding typical height maps obtained by the ST-PSP are reported in Figure \ref{fig_rivulet}(a) and \ref{fig_rivulet}(b), respectively. It should be mentioned that these capillary flows feature maximal free surface elevations of about (a) $0.43\ \rm{mm}$ and (b) $0.66\ \rm{mm}$, significantly smaller than the one for the sessile drop presented in section \ref{subsec3.2}, of about $1.8\ \rm{mm}$. Despite these low height fields, the ST-PSP successfully captures the evolution of their free surface, as illustrated by Figure \ref{fig_rivulet}(c) where the transverse sections taken along the blue and red dashed lines in (a) and (b) are reported in blue and red circles, respectively. For both configurations, the noise amplitude is approximately $13\ \rm{\mu m}$, leading to large $\rm{SNR_{max}}$ values of 33 and 51, respectively, as reported in Table \ref{table_1}. Thus, the method is able to accurately scan a submillimetric capillary object.

Furthermore, theoretical models exist to describe the free surface evolution of a stationary straight rivulet \cite{1966_towell,1995_duffy,1976_bentwich}. Among these models, \citet{1995_duffy} provides the simplest representation of unidirectional flow that accounts for free surface deformation caused by gravity. The authors consider a fluid of constant density $\rho$, surface tension $\gamma$ and dynamic viscosity $\mu$ flowing at a flow rate $Q$ over an inclined substrate forming an angle $\alpha < \pi/2$ with the horizontal and having a contact angle $\theta$ with that surface. They obtained the following evolution for the free surface of the resulting rivulet:

\begin{figure*}
    \centering
    \includegraphics[width=\linewidth]{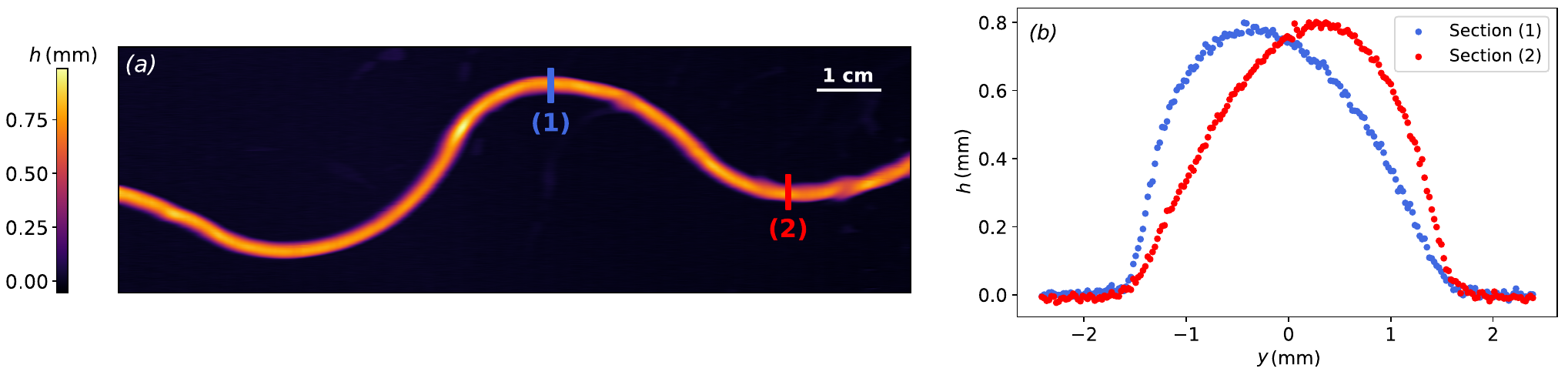}
    \caption{(a) Height distribution obtained using the ST-PSP method for a stationary meander flowing at a rate of $Q = 60\ \rm{mL/min}$ over a plate inclined at $50.3^\circ$ with the horizontal. The colorbar shows the local elevation, in millimeters, while the blue and red dashed lines indicate the positions at which sections have been extracted. The direction of the flow is left to right. (b) Profiles extracted at those positions, the color of the symbols corresponding to the cuts in (a).}
    \label{fig_meander}
\end{figure*}

\begin{equation}
    \label{dm_eq_1}
    \eta(y) = \tan \theta \ell_c \frac{\cosh{\mathrm{Bo}} - \cosh{ \left( y / \ell_c \right) } }{\sinh{\mathrm{Bo}}},
\end{equation}

\noindent where $\ell_c \equiv \sqrt{\gamma / (\rho g \cos \alpha)}$ is the capillary length of the liquid under effective gravity, $g$ the gravitational acceleration, and $\mathrm{Bo}$ is the Bond number which writes $\mathrm{Bo} \equiv \ell / \ell_c$ with $\ell$ the half-width of the rivulet. This Bond number characterizes the balance between gravitational and surface tension effects and it is solution to the non linear conservation equation for the flow rate \cite{1995_duffy}

\begin{equation}
    \label{dm_eq_2}
    \begin{aligned}
    \frac{9 \mu \rho g \cos^2 \alpha}{\gamma^2 \tan^3 \theta \sin \alpha} Q & = 15 \mathrm{Bo} \coth^3{\mathrm{Bo}} - 15 \coth^2{\mathrm{Bo}} \\
    & - 9 \mathrm{Bo} \coth{\mathrm{Bo}} +4.
    \end{aligned}
\end{equation}

\noindent Using equations \eqref{dm_eq_1}-\eqref{dm_eq_2}, only two independent variables — among flow rate $Q$, maximum flow height $\eta_m$ and contact angle $\theta$ — are needed to derive a theoretical free surface profile. In this study, the flow rate $Q$ and the maximum flow height $\eta_m$ were chosen for this purpose. Thus, in Figure \ref{fig_rivulet}(c), the black solid lines correspond to the solutions of equations \eqref{dm_eq_1}-\eqref{dm_eq_2} for the two cases (a) $Q = 8\ \rm{mL/min}$ and $\eta_m \simeq 0.43\ \rm{mm}$, and (b) $Q = 40\ \rm{mL/min}$ and $\eta_m \simeq 0.66\ \rm{mm}$. The experimental profiles and the associated analytical predictions are in very good agreement, which confirms that the ST-PSP is able to successfully capture details of capillary flows at that scale. As such, it is expected that this approach would prove efficient to conduct more systematic studies on the hydrodynamics of such rivulet flows.

\subsection{Stationary meander}\label{subsec3.4}

When the flow rate of the rivulet exceeds a certain threshold, stable meanders begin to form \cite{2006b_le_grand-piteira,2018_grivel}. At even larger flow rates, these meanders can become unsteady \cite{2006b_le_grand-piteira}. However, even in their unsteady state, the meanders can be accurately studied by selecting projection and data acquisition rates that ensure that the meanders remain quasi-stationary during the reconstruction time frame. In this study, a flow rate of $Q=60\ \rm{mL/min}$ has been considered, for which the flow remains quasi stationary. Indeed, in that case, the lateral displacement of the meander in the $y$-direction during the time required for the profilometry reconstruction was found to be, on average, of order $\delta y \simeq 7\ \rm{px} \simeq 0.16\ \rm{mm}$ per reconstruction (\textit{i.e.}, during $\Delta t \simeq 10\ \rm{s}$). Given this minimal displacement relative to the high resolution of the images, the meander can effectively be considered stationary during the data acquisition period. The corresponding ST-PSP reconstruction is presented in Figure \ref{fig_meander}(a). The shape of the meander is accurately captured over a $13\ \rm{cm}$ length with no significant distortions observed, even in regions of high curvature.

Two representative profiles corresponding to the blue and red lines in Figure \ref{fig_meander}(a) are reported in Figure \ref{fig_meander}(b), keeping the same color code for the symbols. They correspond to the free surface elevation at two opposite bends of the meander, with a relatively constant maximal height of about $800\ \mu\rm{m}$. Once more, the value for $\rm{SNR_{max}}$ is large ($\rm{SNR_{max}} = 55$, as reported in Table \ref{table_1}), ensuring that all the flow details are fairly captured. Although the contact lines appear slightly smoothed due to the high pixel density per fringe period ($\lambda_0 = 1.51 \rm{mm}$ corresponds to 64 pixels in the camera's focal plane), the reconstruction still enables the extraction of key physical features, including the meander width, apparent contact angles, and cross-sectional profiles. This allows us, for instance, to observe the asymmetry of the two rivulet sections, that highlights the balance between pinning forces, surface tension and inertia \cite{2006b_le_grand-piteira}. Interestingly, the two plotted bends in Figure \ref{fig_meander}(b) have almost identical widths and are symmetrical with respect to the axis $y=0$, with each section bending due to inertial effects. The contact angles on each side of the meander can also be measured using a linear polynomial fit \cite{2020_quetzeri-santiago}. This reveals a difference $\Delta \theta \simeq 15.1^{\circ}$ between each side of the rivulet for the blue dotted section and $\Delta \theta \simeq 22.7^{\circ}$ for the red dotted section. From this experiment, it can be inferred that the ST-PSP method is not necessarily limited to purely stationary objects, as it could also be applied to situations where the timescale associated with the flow displacements is much longer than the timescale of projection and acquisition.

To the best of our knowledge, meanders have never been experimentally studied using three-dimensional optical methods. Most previous studies relied on photograph acquisition techniques \cite{2006a_le_grand-piteira, 2006b_le_grand-piteira, 2014_couvreur, 2011_daerr} to analyze the geometry of meanders (in terms of width, oscillation amplitude, curvature and wavelength) from a top-down perspective, as well as dynamic meandering instability \cite{2011_daerr, 2008_birnir}. Moreover, no theoretical model has been proposed to rationalize the asymmetry of stationary meanders or the evolution of contact angles along the curvilinear axis of the meander. With the accurate three-dimensional results from the ST-PSP reported here, significant experimental progress is expected to be achieved in order to understand the underlying physical mechanisms of meandering.

\subsection{Braided film}\label{subsec3.5}

When the flow rate is increased sufficiently beyond the meander regime limit, the so-called braided film emerges, characterized by spatial oscillations in the flow width before it eventually stabilizes into a straight rivulet \cite{2004_mertens,2005_mertens,2016_aouad,2017_singh,2018_grivel,2023_brient}. If some attempts have been made to provide a theoretical description of this flow \cite{2005_mertens,2016_aouad}, these analyses make strong assumptions about the free surface geometry of the braided film, or hypothesize a constant contact angle along the flow. However, experimental measurements are currently lacking to confirm or refute these hypotheses.

\begin{figure*}
    \centering
    \includegraphics[width=\linewidth]{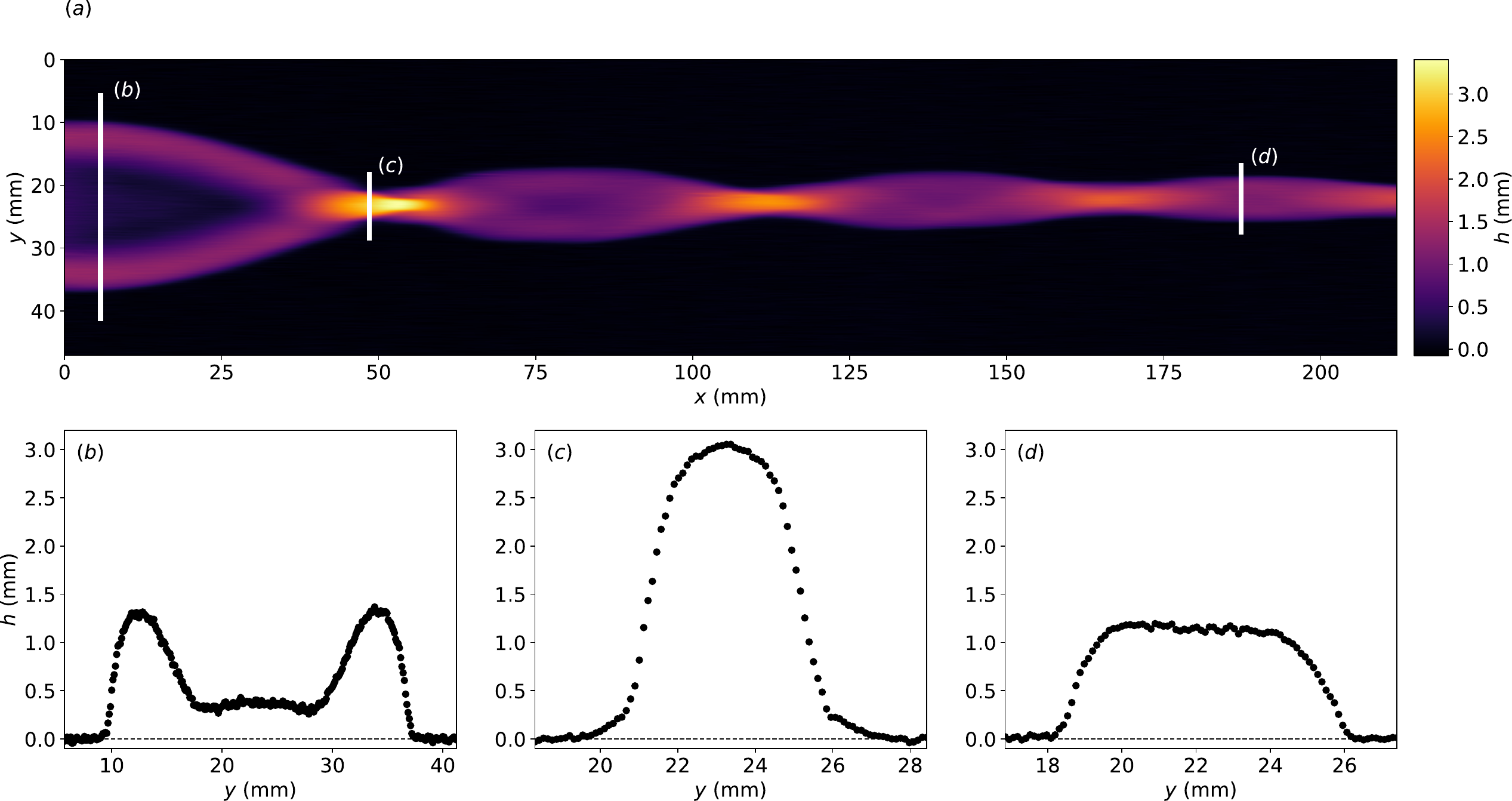}
    \caption{(a) three dimensional reconstruction of a braided film obtained by injecting the liquid normal to a plate inclined at $50.3^\circ$ with the horizontal, at a flow rate of $Q \simeq 520\ \rm{mL/min}$. The flow direction is from left to right. The colorbar shows the local elevation in millimeters, while the white dashed lines indicate the positions at which sections have been extracted. (b)-(d) Section cuts taken at representative locations along the flow: (b) $x \simeq 5.73\ \rm{mm}$, (c) $x \simeq 48.54\ \rm{mm}$ and (d) $x \simeq 187.29\ \rm{mm}$. The black dashed lines correspond to $z=0$.}
    \label{fig_braided_film_1}
\end{figure*}

\begin{table*}
\centering
\caption{Typical noise amplitudes and maximal signal-to-noise ($\rm{SNR_{max}}$) ratios encountered for the different flows investigated in Section \ref{sec3}.}
\label{table_1}
%\resizebox{0.75\linewidth}{!}{%
\begin{tabular}{|c|c|c|c|c|c|}
        \hline
%        & Sessile drop &  Rivulet (8 mL/min)  & Rivulet (40 mL/min) & Stationary meander & Braided film \\
%        & drop & (8 mL/min) & (40 mL/min) & meander & film \\
        & Sessile &  Rivulet  & Rivulet & Stationary & Braided \\
        & drop & (8 mL/min) & (40 mL/min) & meander & film \\
        \hline
        Noise amplitude ($\rm{\mu m}$) & 20 & 13 & 13 & 18 & 32 \\
        %\hline
        $\rm{SNR_{max}}$ & 91 & 33 & 51 & 55 & 106 \\
        \hline
\end{tabular}%
%}
\end{table*}

To produce this distinctive thin film, liquid is injected normal to a plate inclined at $50.3^\circ$ to the horizontal, with a flow rate of $Q \simeq 520\ \rm{mL/min}$. The ST-PSP measurements for the resulting braided film are shown in Figure \ref{fig_braided_film_1}(a). The flow exhibits damped oscillations in its width, as well as significant height variations along the $x$-direction. Near the source region (from $x=0$ to $x=40\ \rm{mm}$), a hydraulic jump is observed, characterized by a thin central region of less than $500\ \mu\rm{m}$ flanked by thicker ropes approximately $1\ \rm{mm}$ thick at the edges, as depicted in Figure \ref{fig_braided_film_1}(b). As $x$ increases, the two outer ropes are drawn toward each other, merging at the first node located around $x \simeq 50\ \rm{mm}$, as shown in Figure \ref{fig_braided_film_1}(c). This convergence results in a pronounced bump, with the free surface elevation reaching approximately $3\ \rm{mm}$. Downstream of this node ($x \in [50, 75]\ \rm{mm}$), the film splits again into two ropes, resembling the structure observed upstream. The contraction process repeats between $x \simeq 75\ \rm{mm}$ and the second node at $x \simeq 110\ \rm{mm}$. This sequence continues, forming four nodes and three braids within the observed region. A representative profile of the final braid is presented in Figure \ref{fig_braided_film_1}(d). A comparison with Figure \ref{fig_braided_film_1}(b) demonstrates significant narrowing of the liquid film along the flow. This behavior highlights the role of viscous dissipation, which competes with inertial and capillary forces \cite{2023_brient}, in attenuating the braided rivulet spatially.

It is interesting to note that such a flow brings several challenges to a profilometry setup, with important variations of the local heights and their gradients, especially at sharp regions close to the nodes, and the need for conducting accurate measurement on a wide area as well as close to the contact line. The profiles represented in \ref{fig_braided_film_1}(b)-(d) show that the ST-PSP is able to overcome these difficulties ($\rm{SNR_{max}}=106$, as reported in Table \ref{table_1}), and could be useful in the aim of conducting detailed studies on these flows. As an illustration of this point, it is clear from an inspection of Figure \ref{fig_braided_film_1}(b) and Figure \ref{fig_braided_film_1}(c) that the contact angle can hardly be considered constant along the flow. From Figure \ref{fig_braided_film_1}(a), it can also be inferred that the contact angle oscillates spatially alongside the width oscillations. These variations in contact angle have not been accounted for in previous theoretical models \cite{2005_mertens}, but their inclusion could be essential for achieving a closer agreement between theoretical predictions and experimental observations.

\section{Conclusion and perspectives}\label{sec4}

In the present study, the spatio-temporal phase shifting profilometry method, initially developed by \citet{2019_ri}, has been adapted to reconstruct the free surface of various capillary flows by dyeing the water solution with a white marker. This method was first calibrated using solid wedges and then tested on a small 3D-printed pyramid to study the influence of the fringe pattern wavelength on the accuracy of the reconstructions. When adapting this method to flowing fluids, dyeing the solution with a white marker is essential to provide sufficient contrast of the projected fringes on the free surface. The minimal concentration of the marker so that the scattering extinction is negligible while minimizing alterations of the fluid's hydrodynamic properties has been determined. It should be emphasized that the present experimental technique is expected to be applicable to most aqueous solutions without major issues, as the dye used here consists of titanium dioxide and chalk particles diluted in a water solution. For more complex or viscous fluids, such as silicon oil, the approach is also expected to work, provided an equivalent method for dyeing the fluid is found. If another dye is used, the protocol described in sections \ref{subsec2.3.3} and \ref{subsec2.4} of the present study should be applied to ensure sufficient light diffusivity at the free surface.

%, enabling accurate reconstruction of the fluid geometry
%was determined to neglect the scattering extinction length and ensure precise measurements of the free surface while minimizing any significant alteration of the fluid's hydrodynamic properties.

In contrast to pre-existing methods, such as FTP, an optimized experimental setup based on the ST-PSP technique enables accurate reconstructions of high surface gradients, contact lines, and apparent contact angles. Four different types of flows were explored: static sessile drop, straight rivulet, stationary meander, and braided film. All reconstructions exhibited large signal-to-noise ratios, thus providing accurate three-dimensional measurements that can be used to analyze qualitatively and quantitatively flow cross-sections and contact angles. Therefore, this three-dimensional profiling method has the ability to enhance the understanding of the physical mechanisms behind complex three-dimensional flows, which often lack comprehensive experimental data to develop analytical models.

\backmatter

%\bmhead{Supplementary information}

\bmhead{Acknowledgements} The authors warmly thank C. Frot for her help in the elaboration of the experimental set-up.

\section*{Declarations}

%Some journals require declarations to be submitted in a standardised format. Please check the Instructions for Authors of the journal to which you are submitting to see if you need to complete this section. If yes, your manuscript must contain the following sections under the heading `Declarations':

%\begin{itemize}
%\item Conflict of interest/Competing interests
%\item Ethics approval and consent to participate
%\item Consent for publication
%\item Data availability 
%\end{itemize}

\bmhead{Funding} This work was partially supported by Agence de l’Innovation de Défense (AID) – via Centre Interdisciplinaire d’Etudes pour la Défense et la Sécurité (CIEDS) – (project 2021 – ICING).

\bmhead{Conflict of interest/Competing interests} The authors report no conflict of interest.

\bmhead{Author contribution}\label{section_code_availability} H. de Miramon and W. Sarlin contributed equally to the present study.

\bmhead{Code availability} The code that supports the findings of the present study is available at \url{https://github.com/heliedemiramon/ST-PSP.git}.

\bmhead{Materials availability} The three dimensional printable model used to manufacture the calibration block depicted in figure \ref{fig_calibration}(a) is available at \url{https://github.com/heliedemiramon/ST-PSP.git}.

\bmhead{Data availability} The data that support the findings of this study are available from the corresponding author, upon reasonable request.

%\noindent
%If any of the sections are not relevant to your manuscript, please include the heading and write `Not applicable' for that section. 

%\bigskip

%\bigskip\noindent
%Springer journals and proceedings: \url{https://www.springer.com/gp/editorial-policies}

%\bigskip\noindent
%Nature Portfolio journals: \url{https://www.nature.com/nature-research/editorial-policies}

%\bigskip\noindent
%\textit{Scientific Reports}: \url{https://www.nature.com/srep/journal-policies/editorial-policies}

%\bigskip\noindent
%BMC journals: \url{https://www.biomedcentral.com/getpublished/editorial-policies}
%\end{flushleft}


\bibliography{bib_profilometry,bib_meanders,bib_rivulets,bib_tio2,bib_braided_films,bib_drops,bib_suspensions}% common bib file

\end{document}